\documentclass[11pt]{article}
\usepackage{amsfonts,latexsym}
\usepackage{amsmath}
\usepackage{amscd}
\usepackage{float,amsmath,amssymb,mathrsfs,bm,multirow,graphics}
\usepackage[dvips]{graphicx}
\usepackage{amsbsy}

\newcommand{\nequation}{\setcounter{equation}{0}}
\renewcommand{\theequation}{\mbox{\arabic{section}.\arabic{equation}}}
\newcommand{\R}{{\Bbb R}}

\newcommand{\C}{{\Bbb C}}

\newcommand{\proofbegin}{\noindent{\quad \it Proof.\,\,}}
\newcommand{\proofend}{\hfill$\Box$\bigskip}

\newtheorem{theorem}{Theorem}[section]
\newtheorem{proposition}[theorem]{Proposition}

\newtheorem{remark}[theorem]{Remark}

\newtheorem{figuretext}{Figure}[section]

\input epsf
\title{\sc Boundary value problems for the stationary axisymmetric Einstein equations: a rotating disk.}
\author{J. Lenells and A. S. Fokas}

\begin{document}

\begin{center}

{\LARGE \sc Boundary value problems for the stationary axisymmetric Einstein equations: a rotating disk
\\} \vspace {7mm}  \noindent

{\large J. Lenells$^{a}$} and {\large A. S. Fokas$^{b}$}

\vskip.7cm

\hskip-.6cm
\begin{tabular}{c}
$\phantom{R^R}^{a}${\small Institut f\"ur Angewandte Mathematik, Leibniz Universit\"at Hannover}\\ 
{\small Welfengarten 1, 30167 Hannover, Germany} \\
{\small E-mail: lenells@ifam.uni-hannover.de} \\
\\
$\phantom{R^R}^{b}${\small Department of Applied Mathematics and Theoretical Physics, University of Cambridge,  }
\\ {\small Cambridge CB3 0WA, United Kingdom} \\
{\small E-mail: t.fokas@damtp.cam.ac.uk} \\
\\
\end{tabular}
\vskip.5cm
\end{center}
\input epsf

\begin{abstract} 
\noindent
The stationary, axisymmetric reduction of the vacuum Einstein equations, the so-called Ernst equation, is an integrable nonlinear PDE in two dimensions. There now exists a general method for analyzing boundary value problems for integrable PDEs, and this method consists of two steps: (a) Construct an integral representation of the solution characterized via a matrix Riemann-Hilbert (RH) problem formulated in the complex $k$-plane, where $k$ denotes the spectral parameter of the associated Lax pair. This representation involves, in general, some unknown boundary values, thus the solution formula is {\it not} yet effective. (b) Characterize the unknown boundary values by analyzing a certain equation called the {\it global relation}. This analysis involves, in general, the solution of a nonlinear problem; however, for certain boundary value problems called linearizable, it is possible to determine the unknown boundary values using only linear operations. Here, we employ the above methodology for the investigation of certain boundary value problems for the elliptic version of the Ernst equation. For this problem, the main novelty is the occurence of the spectral parameter in the form of a square root and this necessitates the introduction of a two-sheeted Riemann surface for the formulation of the relevant RH problem. As a concrete application of the general formalism, it is shown that the particular boundary value problem corresponding to the physically significant case of a rotating disk is a linearizable boundary value problem. In this way the remarkable results of Neugebauer and Meinel are recovered.
 \end{abstract}



\tableofcontents
\section{Introduction}\label{introsec}\nequation
The stationary axisymmetric vacuum Einstein equations can be reduced to a single nonlinear PDE, the so-called Ernst equation. The elliptic version of this equation is given by
\begin{equation}\label{ernst}  
  \frac{f + \bar{f}}{2}\left(f_{\rho\rho} + f_{\zeta\zeta} + \frac{1}{\rho} f_\rho\right) = f_\rho^2 + f_\zeta^2, \qquad \rho > 0, \, \zeta \in \R,
\end{equation}  
where $f(\rho, \zeta)$ is a complex-valued function, called the Ernst potential and bar denotes complex conjugation.

A unified method for analyzing boundary-value problems (BVPs) for linear and integrable nonlinear PDEs in two dimensions was introduced in \cite{F1997} and developed further by several authors \cite{BFS2004, BFS2006, BK2003, BS2003, F2002, F2004, Fbook, FI2004, FIS2005, FLKdV, FM1999, L-FGNLS, MK2006, P2005}. This method consists of two novel steps: (a) Construct an integral representation of the solution by performing the {\it simultaneous} spectral analysis of both parts of the Lax pair (this is to be contrasted with the inverse scattering transform method where one only performs the spectral analysis of the $t$-independent part of the Lax pair). (b) Characterize the unknown boundary values by analyzing the so-called global relation. Step (a) characterizes the solution in terms of a Riemann-Hilbert (RH) problem involving {\it all} boundary values. Thus, in order to obtain an effective solution, it is necessary to implement step (b). The complexity of this step depends on the particular boundary value problem under consideration. For example, for linear evolution PDEs and for some simple BVPs for linear elliptic PDEs, the unknown boundary values can be obtained in closed form. On the other hand, for nonlinear evolution PDEs on the half-line, the unknown boundary values, in general, are characterized through the solution of a nonlinear Volterra integral equation; however, for particular boundary conditions called {\it linearizable}, step (b) can be solved in closed form.

\begin{figure}
\begin{center}
    \includegraphics[width=.4\textwidth]{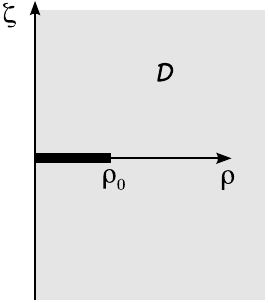} \quad
     \begin{figuretext}\label{diskdomain.pdf}
       The exterior domain $\mathcal{D}$ of a finite disk of radius $\rho_0$.
     \end{figuretext}
 \end{center}
\end{figure}

Here, we implement this methodology to the elliptic version of the Ernst equation (\ref{ernst}). For this problem, the main novelty is the occurence in the Lax pair of the spectral parameter in the form of a square root, which necessitates the introduction of a two-sheeted Riemann surface. A most interesting feature of the Ernst equation is the existence of a large class of linearizable BVPs. Let $\mathcal{D}$ denote the exterior of a finite disk of radius $\rho_0 > 0$ (see Figure \ref{diskdomain.pdf}), i.e.
$$\mathcal{D} = \{(\rho, \zeta) \in \R^2 | \rho > 0\} \backslash \{(\rho, 0) \in \R^2 | 0 < \rho < \rho_0\},$$
and let $f'(\rho, \zeta)$ denote the Ernst potential in coordinates corotating with angular velocity $\Omega$ (see subsection \ref{corotatingsubsec}).
Then the following BVPs are linearizable:
\begin{enumerate}
\item[(A)] {\bf ($f'_\rho = 0$ on a finite disk)}
\begin{itemize}
\item[($i$)] $f$ satisfies (\ref{ernst}) in $\mathcal{D}$.

\item[($ii$)] $f(\rho, \zeta) = \overline{f(\rho, -\zeta)}$ (equatorial symmetry).

\item[($iii$)] $f(\rho, \zeta) \to 1$ as $\rho^2 + \zeta^2 \to \infty$ (asymptotic flatness).

\item[($iv$)] $f_\rho(+0, \zeta) = 0$ for all $\zeta \neq 0$ (regularity on rotation axis).

\item[($v$)] $f'_\rho(\rho, \pm0) = 0$ for $0 < \rho < \rho_0$ (constant Dirichlet boundary conditions on the disk).
\end{itemize}

\item[(B)] {\bf ($f'_\zeta = 0$ on a finite disk)}

The solution $f$ satisfies ($i$)-($iv$) of (A) but ($v$) is replaced with 
\begin{itemize}
\item[($v'$)] $f'_\zeta(\rho, \pm0) = 0$ for $0 < \rho < \rho_0$ (vanishing Neumann boundary conditions on the disk).
\end{itemize}

\item[(C)] {\bf ($f'_\rho = 0$ or $f'_\zeta = 0$ on an infinite disk)}

The solution $f$ satisfies the same BVPs as in (A) and (B), respectively, but the domain $\mathcal{D}$ is replaced with the exterior $\mathcal{D}'$ of an infinite disk stretching from $\rho_0 > 0$ to infinity, i.e.
$$\mathcal{D}' = \{(\rho, \zeta) | \rho > 0\} \backslash \{(\rho, 0) | \rho_0 < \rho\}.$$
In the case of $f_\rho' = 0$, the condition ($iii$) of asymptotic flatness must be replaced with a condition compatible with the constant value of $f'$ on the disk.
\end{enumerate}


The physically significant problem (A) has been studied extensively in the pioneering work of Neugebauer and Meinel \cite{NM1993}-\cite{NM1995} (see also \cite{MAKNP}). It is remarkable that these authors were able to solve this problem {\it without} the guidance of any general method. We can now revisit this problem with the advantage of having at our disposal the general methodology of \cite{F1997} and this, we hope, makes the relevant construction easier to motivate and simpler. 

It was emphasized in \cite{F1997} that the new methodology provides a new approach to solving {\it linear} BVPs. Actually, if a given linear BVP can be solved by the new method, then the corresponding nonlinear problem can be solved following conceptually similar (but analytically more complicated) steps. For this reason we will first investigate the linearized version of equation (\ref{ernst}).

\subsection{Organization of the paper}
In section \ref{axisec} we analyze the axisymmetric Laplace equation, or equivalently the static version of the Ernst equation. In section \ref{ernstsec} we implement step (a) of the new method for the exterior finite disk domain depicted in Figure \ref{diskdomain.pdf}. This yields an expression for the Ernst potential $f$ in terms of the solution of a RH problem which involves certain spectral functions defined in terms of both the Dirichlet and Neumann boundary values on the disk, see Proposition \ref{RHprop}.
In section \ref{eqsec} we analyze the consequences of $f$ being equatorially symmetric and derive the global relation.
In section \ref{linearizablesec} we analyze the global relation for the particular case of the BVP specified in (A). The BVPs formulated in (B) and (C) will be analyzed elsewhere.

\section{The axisymmetric Laplace equation}\label{axisec}\nequation
The linearized version of equation (\ref{ernst}) is the axisymmetric Laplace equation given by
$$f_{\rho \rho} + f_{\zeta \zeta} + \frac{1}{\rho} f_\rho = 0, \qquad \rho > 0, \quad \zeta \in \R.$$
Instead of considering this equation as the linearized approximation of (\ref{ernst}), it is convenient to view it as the exact formulation of (\ref{ernst}) in the case of a static spacetime. A static (as opposed to stationary) spacetime corresponds to a real-valued Ernst potential $f = e^{2U}$, $U \in \R$, and in this case (\ref{ernst}) reduces to the following equation for the real-valued function $U(\rho, \zeta)$:
\begin{subequations}
\begin{equation}\label{axilaplace}  
 U_{\rho \rho} + U_{\zeta \zeta} +  \frac{1}{\rho} U_\rho = 0, \qquad \rho > 0, \quad \zeta \in \R.
\end{equation}  
The boundary conditions on the rotation axis and at infinity are
\begin{align}\label{axiUrotationaxis}  
  & U_\rho(+0, \zeta) = 0  \quad \hbox{for} \quad \zeta \neq 0,
  	\\ \label{axiUflatness}
 & U(\rho, \zeta) \to 0 \quad \hbox{as} \quad \rho^2 + \zeta^2 \to \infty.
  \end{align}
\end{subequations}

\subsection{Lax pair}
It is convenient to work with a complex variable $z = \rho + i\zeta$ and write $U(z)$ for $U(\rho, \zeta)$. For convenience of notation we will suppress the dependence on $\bar{z}$, i.e. in general $f(z)$ will denote a function depending on both $z$ and $\bar{z}$.

Equation (\ref{axilaplace}) admits the following Lax pair formulation for the scalar function $\phi(z, k)$:
\begin{equation}\label{axilax}  
\begin{cases}
\phi_z(z, k) = \lambda U_z(z),	\\
\phi_{\bar{z}} (z, k) = \frac{1}{\lambda} U_{\bar{z}}(z),
\end{cases}
\end{equation}
where $\lambda = \lambda(z, k)$ is defined by
\begin{equation}\label{lambdadef}
  \lambda = \left(\frac{k - i \bar{z}}{k + i z}\right)^{1/2},
\end{equation}
and $k \in \C$ is a spectral parameter. We write (\ref{axilax}) in the differential form
\begin{equation}\label{axilaxdiffform}  
  d\phi = W,
\end{equation}  
where $W = W(z, k)$ is the one-form
\begin{align} \label{axiWdef}
W = & \lambda U_z dz + \frac{1}{\lambda} U_{\bar{z}} d\bar{z}
	\\ \nonumber
=& \frac{1}{2}\left[\left(\frac{1}{\lambda} + \lambda\right)U_\rho + i\left(\frac{1}{\lambda} - \lambda\right)U_\zeta \right]d \rho
+ \frac{1}{2}\left[\left(\frac{1}{\lambda} + \lambda\right)U_\zeta - i\left(\frac{1}{\lambda} - \lambda\right)U_\rho \right]d\zeta.
\end{align}

\subsection{A two-sheeted Riemann surface}
According to the methodology introduced in \cite{F2002}, the solution of the so-called direct problem involves the construction of a solution $\phi(z, k)$ of equation (\ref{axilaxdiffform}), which is bounded for all $k \in \C$. For a polygonal domain, this can be achieved by integrating (\ref{axilaxdiffform}) along a contour starting at a corner of the domain. In particular, the corner $i\infty$ yields the solution
\begin{equation}\label{phidef1}
  \phi(z, k) = \int_{i \infty}^z W(z', k).
\end{equation}
However, the definition (\ref{lambdadef}) of $\lambda$ involves a square root, hence the value of the right-hand side of (\ref{phidef1}) will depend on the path of integration since the path affects the choice of the branch of the relevant square root. In order for the right-hand side of (\ref{phidef1}) to be {\it independent} of the path of integration (so that $\phi$ is well-defined), we introduce the following genus $0$ Riemann surface $\mathcal{S}_z$: For each value of $z = \rho + i\zeta$, $\mathcal{S}_z$ consists of the set of points $(\lambda, k) \in \C^2$ such that
$$\lambda^2 = \frac{k - i \bar{z}}{k + i z}.$$
We introduce a branch cut in the complex $k$-plane from $-iz$ to $i\bar{z}$ and, for $k \in \C$, we let $k^+$ and $k^-$ denote the corresponding points on the upper and lower sheet of $\mathcal{S}_z$, respectively. By definition, the upper (lower) sheet is characterized by $\lambda \to 1$ ($\lambda \to -1$) as $k \to \infty$. We compactify $\mathcal{S}_z$ by adding the two points $\infty^+$ and $\infty^-$. For each $z$, $\phi(z, \cdot)$ is a map $\mathcal{S}_z \to \C$. 
The definition (\ref{phidef1}) of $\phi(z, k)$ is made precise by choosing $k$ to lie on the lower sheet initially when $z' = i\infty$. This picks one of the two branches of the square root defining $\lambda(z', k)$. We then use analytic continuation to follow this branch throughout the integration. This defines the integrand unambiguously. 
At the end of the integration, i.e. when $z'$ approaches $z$, $k$ will lie either on the upper or on the lower sheet of $\mathcal{S}_z$; we denote the corresponding two values of $\phi$ by $\phi(z, k^+)$ and $\phi(z, k^-)$, respectively. 

\subsection{The direct problem}
\begin{figure}
\begin{center}
    \includegraphics[width=.26\textwidth]{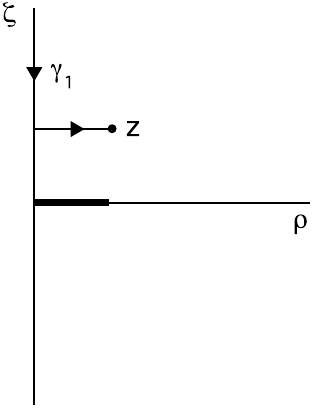} \quad
    \includegraphics[width=.405\textwidth]{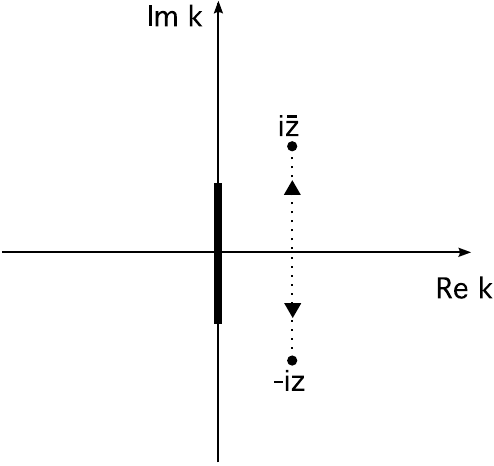} 
    	\\
    \includegraphics[width=.26\textwidth]{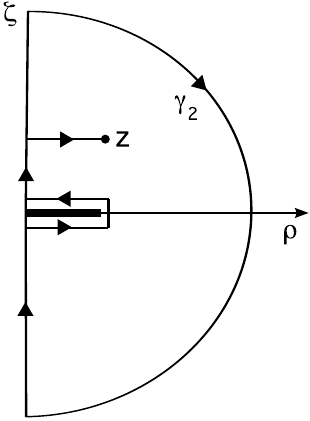} \quad
    \includegraphics[width=.405\textwidth]{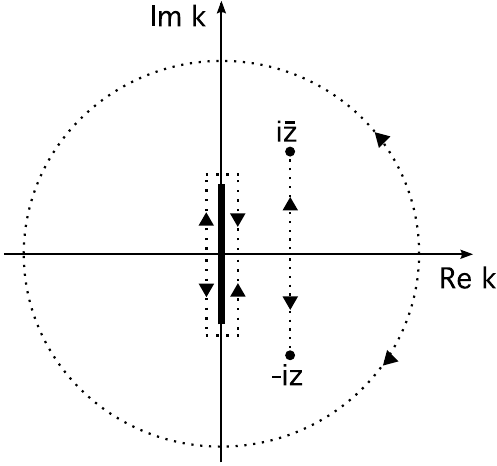} 
     \begin{figuretext}\label{gamma12contours.pdf}
       The graphs on the left show the integration contours $\gamma_1$ and $\gamma_2$ used to define $\phi(z, k^-)$ and $\phi(z, k^+)$ for $\zeta = \text{\upshape Im}\, z > 0$. The graphs on the right illustrate how the endpoints of the branch cut in the complex $k$-plane move as these contours are traversed. For $z'$ on the $\zeta$-axis the Riemann surface degenerates and the branch cut disappears.
  \end{figuretext}
 \end{center}
\end{figure}

In order to define a function $\phi(z, k)$ for all $k$ on $\mathcal{S}_z$, starting fromÊ $z' = i\infty$, we integrate with respect to the contours $\gamma_1$ and $\gamma_2$, see graphs on the left of Figure \ref{gamma12contours.pdf}. It turns out that for $\zeta > 0$, the function $\phi$ defined with respect to $\gamma_1$ and $\gamma_2$ lives on the lower and upper sheet of $\mathcal{S}_z$, respectively, i.e.
\begin{align}\label{axiphipositivezeta}
\phi(z, k^-) = \int_{\gamma_1} W(z', k), \qquad \phi(z, k^+) = \int_{\gamma_2} W(z', k).
\end{align}
Indeed, as $z'$ moves along $\gamma_1$ and $\gamma_2$, the endpoints of the branch cut in the $k$-plane move along the dotted curves of Figure \ref{gamma12contours.pdf}. 
Note that for $z = i \zeta$ on the $\zeta$-axis, the Riemann surface $\mathcal{S}_z$ degenerates and consists of two disjoint copies of the complex $k$-plane; we have $\lambda = 1$ for all $k^+$ on the upper sheet and $\lambda = -1$ for all $k^-$ on the lower sheet. Therefore, when integrating along the $\zeta$-axis, the branch cut disappears.

For the movement of the branch cut we note that the branch points for $z' = \rho' + i\zeta'$ occur at $(\zeta' - i\rho', \zeta' + i \rho')$. Any fixed $k$ on the lower sheet will remain on the lower sheet throughout an integration along $\gamma_1$, because $k$ never crosses the branch cut.\footnote{The case when $k$ lies directly on the segment $[-iz, i\bar{z}]$ can be handled by slightly deforming the contour $\gamma_1$.} On the other hand, as the integration along the large semicircle of $\gamma_2$ is performed, the branch cut sweeps across the whole $k$-plane, so that $k$ moves from the lower to the upper sheet. Throughout the rest of the integration along $\gamma_2$, $k$ remains on the upper sheet. 

Since $W(z', k)$ is analytic in $k$ away from the branch cut determined by $z'$, equation (\ref{axiphipositivezeta}) defines $\phi(z, \cdot)$ as an analytic function $\mathcal{S}_z \to \C$ whenever $k$ does not lie on one of the dotted curves in Figure \ref{gamma12contours.pdf} associated with the movements of the branch cuts. Thus there are only two segments across which $\phi$ may have singularities:

\begin{enumerate}
\item The branch cut $[-iz, i\bar{z}]$ of $\mathcal{S}_z$. The values of $\phi(z, k^-)$ to the left (right) of this cut glue together smoothly with the values of $\phi(z, k^+)$ to the right (left). Thus, $\phi$ is analytic as a map $\mathcal{S}_z \to \C$ near this cut.

\item The contour $\Gamma$ defined as the segment in the upper sheet of $\mathcal{S}_z$ lying above $[-i \rho_0, i \rho_0]$. This segment, which arises from integration along the disk, is present only for the contour $\gamma_2$ (hence it lies only in the upper sheet). We will show in section \ref{linearinversesubsec} that $\phi(z, \cdot)$ {\it does} have a jump across $\Gamma$ and we will express this jump in terms of the boundary values of $U$ on the disk. 
\end{enumerate}

\begin{remark}\upshape
1. The Lax pair (\ref{axilax}) has singularities at the two branch points $k = -iz$ and $k = i\bar{z}$ at which $\lambda = \infty$ and $\lambda = 0$, respectively. However, as we verify in detail in appendix \ref{singapp}, the eigenfunction $\phi(z, k)$ is still analytic near these points. 

2. The boundary condition (\ref{axiUrotationaxis}) ensures that the integration along the $\zeta$-axis, which involves the degenerate Riemann surfaces, is compatible with the integration involving nondegenerate Riemann surfaces. 
\end{remark}

\subsection{The inverse problem}\label{linearinversesubsec}
\begin{figure}
\begin{center}
    \includegraphics[width=.65\textwidth]{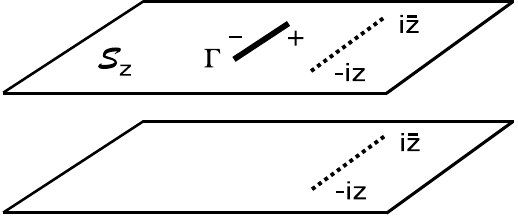} 
         \begin{figuretext}\label{Gamma.pdf}
       The contour $\Gamma$ and the Riemann surface $\mathcal{S}_z$ used for the RH problem.
     \end{figuretext}
     \end{center}
\end{figure}

Equations (\ref{axiphipositivezeta}) provide the solution of the direct problem, namely they express $\phi$ in terms of $U$ for all $k \in \mathcal{S}_z$.
In order to solve the inverse problem, we must find an alternative representation for $\phi$, namely we must express $\phi$ in terms of an appropriate spectral function. This can be achieved by formulating a RH problem on $\mathcal{S}_z$ (see Figure \ref{Gamma.pdf}). It was shown in the previous subsection that 
\begin{subequations}\label{axiRH}
\begin{equation}\label{axiRH1}
  \text{$\phi(z, k)$ is analytic for $k \in \mathcal{S}_z \backslash \Gamma$.} 
\end{equation}  
Thus, in order to formulate a RH problem for $\phi$ we must compute the `jump' of $\phi$ across $\Gamma$ in terms of the boundary values of $U$ on the disk. Introducing the notations
$\phi^+$ and $\phi^-$ for the values of $\phi$ to the right and left of $\Gamma$, i.e.,
\begin{align*} 
 \phi^+(z, k) := \phi\left(z, (+0 + ik_2)^+\right), \qquad  \phi^-(z, k) := \phi\left(z, (-0 + ik_2)^+\right), \qquad -\rho_0 \leq k_2 \leq \rho_0,
\end{align*}
and denoting the jump by $D(k)$, i.e.,
\begin{equation}\label{axiRH3}  
  D(k) = \phi^+(z, k) - \phi^-(z, k), \qquad k \in \Gamma,
\end{equation}
\end{subequations}
we will show that
\begin{align}\label{axiDexpression}
  D(k) = 2 \int_{k_2}^{\rho_0}  \frac{ -ik_2 (U_\rho^+  - U_\rho^-)  + \rho(U_\zeta^+  - U_\zeta^-) }{\sqrt{\rho^2 - k_2^2}} d\rho, \qquad k = ik_2 \in \Gamma, \quad k_2 > 0,
\end{align}
where $U_\rho^\pm$ and $U_\zeta^\pm$ denote the Dirichlet and Neumann boundary values of $U$ on the disk, respectively, i.e.
$$U_\rho^\pm := U_\rho(\rho \pm i0), \qquad U_\zeta^\pm := U_\zeta(\rho \pm i0).$$

In order to prove (\ref{axiDexpression}) we need to evaluate $W(z, k)$ for $k$ near the branch cut $[\zeta - i\rho, \zeta + i\rho]$. Suppose $k = (\zeta \pm 0 + ik_2)^+$, $k_2 > 0$, lies on the upper sheet. Then the expression (\ref{lambdadef}) for $\lambda$ yields
$$\lambda|_{k = (\zeta + 0 + ik_2)^+} = \begin{cases}
\sqrt{\frac{k_2 - \rho}{k_2 + \rho}}, \qquad \rho < k_2, \\
-i\sqrt{\frac{\rho  - k_2}{\rho + k_2}}, \qquad \rho > k_2.
\end{cases}$$
Hence
\begin{subequations}\label{lambdacombinations}
\begin{align}
\frac{1}{\lambda} + \lambda\biggl|_{k = (\zeta + 0 + ik_2)^+}
= \begin{cases}
\frac{2k_2}{\sqrt{k_2^2 - \rho^2}}, \qquad \rho < k_2, \\
\frac{2ik_2}{\sqrt{\rho^2 - k_2^2}}, \qquad \rho > k_2, 
\end{cases}
	\\
\frac{1}{\lambda} - \lambda\biggl|_{k = (\zeta + 0 + ik_2)^+}
= \begin{cases}
\frac{2\rho}{\sqrt{k_2^2 - \rho^2}}, \qquad \rho < k_2, \\
\frac{2i\rho}{\sqrt{\rho^2 - k_2^2}}, \qquad \rho > k_2.
\end{cases}
\end{align}
Similarly,
\begin{align}
\frac{1}{\lambda} + \lambda\biggl|_{k = (\zeta - 0 + ik_2)^+}
= \begin{cases}
\frac{2k_2}{\sqrt{k_2^2 - \rho^2}}, \qquad \rho < k_2, \\
\frac{-2ik_2}{\sqrt{\rho^2 - k_2^2}}, \qquad \rho > k_2, 
\end{cases}
	\\
\frac{1}{\lambda} - \lambda\biggl|_{k = (\zeta - 0 + ik_2)^+}
= \begin{cases}
\frac{2\rho}{\sqrt{k_2^2 - \rho^2}}, \qquad \rho < k_2, \\
\frac{-2i\rho}{\sqrt{\rho^2 - k_2^2}}, \qquad \rho > k_2.
\end{cases}
\end{align}
\end{subequations}
Letting\footnote{This notation turns out to be convenient and will be used frequently for functions of $z$ and $k$: a superscript $R$ or $L$ on a function evaluated at a point $(z, k^+)$ with $k^+$ lying on the branch cut $[\zeta - i\rho, \zeta + i\rho]$ determined by $z = \rho + i \zeta$, means that the value of $k^+$ should be shifted infinitesimally to the right or left of the branch cut on the upper sheet before evaluation.}
$$W^R(z, k^+) := W\left(z, (\zeta + 0 + ik_2)^+\right), \qquad
W^L(z, k^+) := W\left(z, (\zeta - 0 + ik_2)^+\right),$$
the definition (\ref{axiWdef}) of $W$ implies
\begin{equation}\label{axiWrightorabovecut}
W^R(z, k^+)
= \begin{cases}
\frac{1}{i\sqrt{k_2^2 - \rho^2}}\left[\left(ik_2 U_\rho - \rho U_\zeta \right)d \rho
+ \left(i k_2 U_\zeta + \rho U_\rho \right)d\zeta\right], \qquad \rho < k_2, \\
\frac{1}{\sqrt{\rho^2 - k_2^2}}\left[\left(ik_2 U_\rho - \rho U_\zeta \right)d \rho
+ \left(i k_2 U_\zeta + \rho U_\rho \right)d\zeta\right], \qquad \rho > k_2
\end{cases}
\end{equation}
and
\begin{equation}\label{axiWLR1}
W^L(z, k^+)
= \begin{cases}
W^R(z, k^+), \qquad \rho < k_2, \\
-W^R(z, k^+), \qquad \rho > k_2.
\end{cases}
\end{equation}
The jump $D(k)$ of $\phi$ across $\Gamma$ can be determined by considering the integral $\int_{\gamma_2} W$ which defines $\phi(z, k^+)$ for $k = \pm 2\epsilon + ik_2$ and taking $\epsilon \to 0$. We choose the integration contour $\gamma_2$ so that when it passes along the disk, it lies a distance $\epsilon$ above or below the disk (so that $k$ remains on upper sheet throughout the integration along the disk). The integrals along the subcontours of $\gamma_2$ which are not along the disk are continuous across $\Gamma$. Hence
\begin{align}\label{axidiskintegrals}
  D(k) =  \int_{[0, \rho_0]} [&W^{R}(\rho - i0, k^+) - W^{R}(\rho + i0, k^+)
  	\\ \nonumber
& - W^{L}(\rho - i0, k^+) + W^{L}(\rho + i0, k^+)], \qquad k \in \Gamma.
\end{align}
In view of (\ref{axiWLR1}), this can be rewritten as
\begin{align} \label{axiDeltaphizetapositive}
  D(k) = & 2 \int_{[k_2, \rho_0]} [W^{R}(\rho - i0, k^+) - W^{R}(\rho + i0, k^+)].
\end{align}
Using in (\ref{axiDeltaphizetapositive}) the expression for $W^R$ given by (\ref{axiWrightorabovecut}), we find (\ref{axiDexpression}).

The values of $D(k)$ for $k = ik_2 \in \Gamma$, $k_2 < 0$, can be obtained from (\ref{axiDexpression}) by symmetry. Indeed, the relation
\begin{equation*}
  \lambda(z, k^\pm) = \frac{1}{\overline{\lambda(z, \bar{k}^\pm)}},
\end{equation*}
implies that
\begin{equation} \label{axiWbarsymmetry}
  W(z, k^\pm) = \overline{W(z, \bar{k}^\pm)}.
\end{equation}
It follows that $\phi$ admits the symmetry
\begin{equation}\label{phiconjugationsymmetry}  
  \phi(z, k^\pm) = \overline{\phi(z, \bar{k}^\pm)},
\end{equation}
and so
$$ \overline{D(\bar{k})} = D(k), \qquad k \in \Gamma.$$

Equations (\ref{axiRH}) constitute a RH problem for $\phi(z, k)$ with jump across $\Gamma$. In order for the solution of this RH problem to be unique we need to supplement equations (\ref{axiRH}) with a normalization condition for $\phi(z,k)$. 
The value of $\phi(z, k^+)$ is defined by integration along $\gamma_2$ according to (\ref{axiphipositivezeta}). The integration along the semicircle vanishes, so that $\phi(-i\infty, k^+) =0$. Therefore, in the limit $k \to \infty$, we find
\begin{equation}\label{axiUfromlimit}
  \lim_{k \to \infty} \phi(z, k^+) = \lim_{k \to \infty} \int_{-i\infty}^z W(z, k^+) = \int_{-i\infty}^z dU = U(z).
\end{equation}  
Similarly,
\begin{equation}\label{axiUlimit2}
  \lim_{k \to \infty} \phi(z, k^-) = -U(z).
\end{equation}  
In particular,
\begin{equation}\label{axiRH2}
  \phi(z, \infty^+) = -\phi(z, \infty^-).
\end{equation}  
Equation (\ref{axiRH2}) provides the required normalization condition which ensures uniqueness of the solution of the RH problem.\footnote{If $\phi_1(z,k)$ and $\phi_2(z, k)$ are two solutions of (\ref{axiRH}) satisfying (\ref{axiRH2}), then $\phi_1 - \phi_2$ is analytic everywhere on $\mathcal{S}_z$. Hence $\phi_1 - \phi_2$ is a constant. Since $(\phi_1 - \phi_2)(z, \infty^+) = -(\phi_1 - \phi_2)(z, \infty^-)$, this constant is zero.}

The unique solution of the RH problem (\ref{axiRH}) together with the normalization condition (\ref{axiRH2}) is given by
\begin{equation}\label{axiRHsolution}
  \phi(z, k) = -\frac{1}{4\pi i}\int_\Gamma D(k')\left(\frac{\lambda(z, k)(k + iz)}{\lambda(z, k')(k' + iz)} + 1\right) \frac{dk'}{k' - k}.
\end{equation}
Indeed, the right-hand side of (\ref{axiRHsolution}) is an analytic function of $(\lambda, k) \in \mathcal{S}_z$ for $k \notin [-i\rho_0, i\rho_0]$. The standard Plemelj formulas imply that this function satisfies the jump condition (\ref{axiRH3}) across $\Gamma$ in the upper sheet, whereas it does not jump on the lower sheet. Finally, the condition (\ref{axiRH2}) is a consequence of the relations $\lambda(z, \infty^+) = - \lambda(z, \infty^-) = 1$.

Equations (\ref{axiphipositivezeta}) express $\phi$ in terms of $U$ (the solution of the direct problem), whereas equation (\ref{axiRHsolution}) expresses $\phi$ in terms of $D(k)$ (the solution of the inverse problem). Using these two different representations of $\phi$ it is straightforward to compute $U$ in terms of $D(k)$: 
Substituting the representation (\ref{axiRHsolution}) into (\ref{axiUfromlimit}), we find
\begin{equation}\label{preaxiUrep}
  U(z) = \frac{1}{4\pi i}\int_\Gamma \frac{D(k)}{\lambda(z, k)(k + iz)} dk.
  \end{equation}
This can be written as
\begin{equation}\label{axiUrepresentation}
  U(z) = -\frac{1}{4\pi i}\int_{-i\rho_0}^{i\rho_0} \frac{D(k)}{\sqrt{(k - \zeta)^2 + \rho^2}} dk,
\end{equation}  
where the branch with positive real part is chosen for the square root.

\begin{remark}\upshape
In the above discussion we assumed $\zeta > 0$. The case of $\zeta < 0$ is similar.
\end{remark}

\subsection{The spectral functions}
The expression (\ref{axiDexpression}) for $D(k)$ involves both the Dirichlet and the Neumann boundary values. However, for a well-posed problem only one of these boundary values is specified. In sections \ref{lineareqsymmsubsec} and \ref{linearglobalsubsec} we will use equatorial symmetry and the global relation to express $D(k)$ in terms of either the Dirichlet or the Neumann boundary values. In this connection, we introduce the following definitions: For $k = ik_2$ with $0 \leq k_2 \leq \rho_0$, the spectral functions $R(k)$ and $S(k)$ are defined by
\begin{subequations}\label{axiRSdef} 
\begin{equation}\label{axiRdef}  
  R(k) := \int_{[0, k_2]} W^{R}(\rho + i0, k^+) = \int_0^{k_2} \frac{ik_2 U_\rho^+ - \rho U_\zeta^+ }{i\sqrt{k_2^2 - \rho^2}} d \rho
\end{equation}
and
\begin{equation}\label{axiSdef}  
  S(k) := \int_{[k_2, \rho_0]} W^{R}(\rho + i0, k^+) = \int_{k_2}^{\rho_0} \frac{ik_2 U_\rho^+ - \rho U_\zeta^+}{\sqrt{\rho^2 - k_2^2}} d \rho.
\end{equation}
\end{subequations}
Note that the functions  $\text{Re}(R)$ and $\text{Re}(S)$ are defined in terms of the Dirichlet and Neumann boundary values of $U$ on the disk, respectively.

\subsection{Equatorial symmetry}\label{lineareqsymmsubsec}
For a variety of physically significant BVPs, we expect the spacetime metric to be symmetric with respect to the equatorial plane. In terms of the Ernst potential this means that $\overline{f(\bar{z})} = f(z)$, so that in the static case $U(z) = U(\bar{z})$. 

\begin{proposition}\label{axiWdiskprop}
Assume that $U(z)$ is equatorially symmetric, i.e. $U(z) = U(\bar{z})$. Then the spectral functions $D(k)$ and $S(k)$, defined by (\ref{axiDexpression}) and (\ref{axiSdef}), are related by
\begin{align}\label{axiDNeumann}
  D(k) = -4\text{\upshape Re}(S(k)).
\end{align}
Furthermore, for $z$ on the disk, $z = \rho \pm i0$, 
\begin{equation}\label{axiWsymm1}
  W^{L}(\rho - i 0, k^+) = \overline{W^{R}(\rho + i0, k^+)}, \qquad k \in \Gamma.
\end{equation}
\end{proposition}
\proofbegin
The symmetry $U(z) = U(\bar{z})$ implies that
\begin{equation}\label{UrhoUzetasymm}  
  U_\rho(z) = U_\rho(\bar{z}),\qquad U_\zeta(z) = -U_\zeta(\bar{z}).
\end{equation}
In particular, the boundary values above and below the disk are related by $U_\rho^+ = U_\rho^-$ and $U_\zeta^+ = - U_\zeta^-$. Using these relations in (\ref{axiDexpression}) and comparing the resulting equation with (\ref{axiSdef}), we find (\ref{axiDNeumann}).

Let $W = W_1 d\rho + W_2 d\zeta$. By the definition (\ref{axiWdef}) of $W$, we have
$$W_1(\bar{z}, k^+) = \frac{1}{2}\left[\left(\frac{1}{\lambda} + \lambda\right)U_\rho(\bar{z}) + i\left(\frac{1}{\lambda} - \lambda\right)U_\zeta(\bar{z}) \right]_{\lambda = \lambda(\bar{z}, k^+)}.$$
Using (\ref{UrhoUzetasymm}) together with the identity
\begin{align}  \label{lambdazzbarsymmetry}
  -\frac{1}{\lambda(\bar{z}, k^+)} = \lambda(z, -k^-),
\end{align}
we find
$$W_1(\bar{z}, k^+) = \frac{1}{2}\left[\left(-\lambda - \frac{1}{\lambda}\right)U_\rho(z) - i\left(-\lambda + \frac{1}{\lambda}\right)U_\zeta(z) \right]_{\lambda = \lambda(z, -k^-)}.$$
But the right-hand side of this equation is precisely $-W_1(z, -k^-)$. 
The relation
$$\lambda(z, k^+) = -\lambda(z, k^-),$$
implies that
\begin{equation}\label{axiWlift1}
  W(z, k^+) = -W(z, k^-).
\end{equation}
Thus, utilizing the symmetries (\ref{axiWbarsymmetry}) and (\ref{axiWlift1}), we find
$$W_1(\bar{z}, k^+) = -W_1(z, -k^-) = W_1(z, -k^+) = \overline{W_1(z, -\bar{k}^+)}.$$
The identity (\ref{axiWsymm1}) now follows by taking $z = \rho + i0$ and observing that if $k^+ = -0 + ik_2$ lies just to the left of $\Gamma$, then $-\bar{k}^+ = 0 + ik_2$ lies just to the right of $\Gamma$.
\proofend

Equations (\ref{axiSdef}) and (\ref{axiDNeumann}) express the jump $D(k)$ in terms of the Neumann boundary values. Thus, inserting equation (\ref{axiDNeumann}) into the right-hand side of (\ref{axiUrepresentation}), we immediately obtain the solution of an equatorially symmetric Neumann boundary value problem. In order to express $D(k)$ in terms of the Dirichlet boundary values, we will use the global relation.

\subsection{The global relation}\label{linearglobalsubsec}
\begin{figure}
\begin{center}
    \includegraphics[height=.37\textwidth]{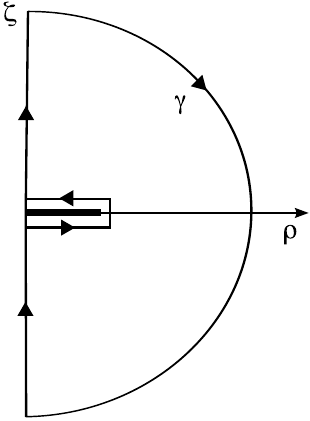} \quad
    \includegraphics[height =.4\textwidth]{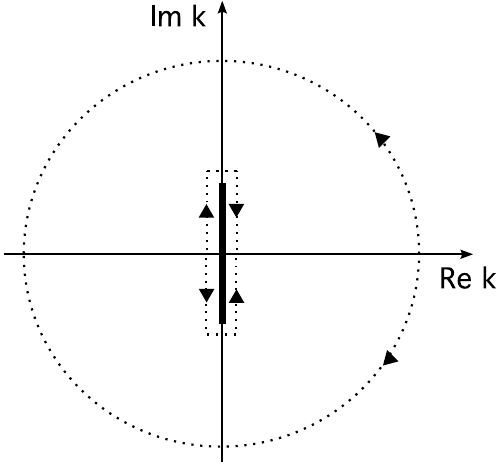} 
     \begin{figuretext}\label{globalcontour.pdf}
       The integration contour encircling all of physical space used for the global relation and the associated movement of the endpoints of the branch cut.
     \end{figuretext}
     \end{center}
\end{figure}

The global relation is an algebraic equation satisfied by the spectral functions. It expresses the fact that the boundary values are related and cannot be independently prescribed. For our problem the global relation can be derived from the equation
\begin{equation}\label{globalrelationgeneral}
  \int_{\gamma} W = 0,
\end{equation}
where $\gamma$ is the contour encircling the physical space depicted in Figure \ref{globalcontour.pdf}.
The identity (\ref{globalrelationgeneral}) is obtained by applying Stokes' theorem together with the fact that $dW = 0$ in view of (\ref{axilaxdiffform}). 

Since $W$ depends on $k$, (\ref{globalrelationgeneral}) is a family of relations parametrized by $k$. 
We choose $k = \epsilon + ik_2$ and specify the integration contour so that when it passes along the disk, it lies a distance $2\epsilon$ above or below the disk (so that $k$ changes sheets during the integration along the disk). Equation (\ref{globalrelationgeneral}) then reads
\begin{align}\label{axiglobalintegrals}
  & \int_{[-i\infty, 0]} W(z,k^+) 
  + \int_{[0, \rho_0]} W^R(\rho - i0,k^+)
  	\\ \nonumber
  & + \int_{[\rho_0, 0]} W^L(\rho + i0,k^-)
  + \int_{[0, i\infty]} W(z,k^-) = 0.
\end{align}
We claim that the two integrals along the $\zeta$-axis can be simplified as follows:
\begin{equation}  \label{axiaxisintegrals}
  \int_{[-i\infty, 0]} W(z,k^+) = U(-i0), \qquad \int_{[0, i\infty]} W(z,k^-) = U(+i0).
\end{equation}
Indeed, for $z = i\zeta$ the form of (\ref{axilaxdiffform}) is particularly simple since $\lambda = 1$ ($\lambda = -1$) for all $k$ on the upper (lower) sheet. We find, for any $\zeta \in \R$ and $k \in \C$,
\begin{align*}
   W(i \zeta, k^+) = dU, \qquad W(i \zeta, k^-) = - dU.
\end{align*}
Together with the initial conditions $\phi(i\infty, k^-) = \phi(-i\infty, k^+) = 0$, this shows (\ref{axiaxisintegrals}).
This yields the following expression for (\ref{globalrelationgeneral}):
\begin{align}\label{axiglobalrelation}
  U(-i0) 
  + \int_{[0, \rho_0]} W^R(\rho - i0,k^+)
  + \int_{[\rho_0, 0]} W^L(\rho + i0,k^-)
  + U(+i0) = 0.
\end{align}

In the presence of the equatorial symmetry, equation (\ref{axiglobalrelation}) can be simplified further. Indeed, in view of Proposition \ref{axiWdiskprop} and the symmetries (\ref{axiWLR1}) and (\ref{axiWlift1}), we have
\begin{align}
  \int_{[0, \rho_0]} W^R(\rho - i0,k^+) = \int_{[0, k_2]} W^L(\rho - i0,k^+) - \int_{[k_2, \rho_0]} W^L(\rho - i0,k^+)
  = \overline{R(k)} - \overline{S(k)},
  	\\
   \int_{[\rho_0, 0]} W^L(\rho + i0,k^-) = \int_{[\rho_0, k_2]} W^R(\rho + i0,k^+) - \int_{[k_2, 0]} W^R(\rho + i0,k^+) 
   =  -S(k) + R(k).
\end{align}  
Substituting these expressions into (\ref{axiglobalrelation}), we find the following result.

\begin{proposition}
Suppose that $U(z)$ obeys the equatorial symmetry $U(z) = U(\bar{z})$.
Then the spectral functions $R(k)$ and $S(k)$ defined by (\ref{axiRSdef}) satisfy the global relation
\begin{equation}\label{axiglobalRS}
 U(+i0) + \text{\upshape Re}(R(k)) = \text{\upshape Re}(S(k)).
\end{equation}
\end{proposition}

Since $\text{Re}(R)$ and $\text{Re}(S)$ involve only the Dirichlet and Neumann boundary values of $U$ on the disk, equation (\ref{axiglobalRS}) defines the Dirichlet to Neumann map.
In particular, equations (\ref{axiDNeumann}) and (\ref{axiglobalRS}) yield
\begin{equation}\label{axiDDirichlet}
  D(k) = -4\left(U(+i0) + \text{Re}(R(k))\right),
\end{equation}
which expresses $D$ in terms of only the Dirichlet boundary values.

We summarize our discussion of the linear problem in the following theorem.

\begin{theorem}\label{axith}
Let $U(\rho, \zeta)$ be a real-valued solution of the axisymmetric Laplace equation (\ref{axilaplace}) in the exterior disk domain $\mathcal{D}$ satisfying the boundary conditions (\ref{axiUrotationaxis}) and (\ref{axiUflatness}). Suppose that $U$ is equatorially symmetric, $U(\rho, \zeta) = U(\rho, -\zeta)$.
Then $U(\rho, \zeta)$ admits the integral representation
\begin{equation}\label{axiUfinal}
U(\rho, \zeta) = -\frac{1}{4\pi i}\int_{-i\rho_0}^{i\rho_0} \frac{D(k)}{\sqrt{(k - \zeta)^2 + \rho^2}} dk, \qquad (\rho, \zeta) \in \mathcal{D},
\end{equation}
where the branch with positive real part is chosen for the square root, and the function $D(k)$ is given in terms of the Dirichlet and Neumann boundary values of $U$ on the disk by the following expressions respectively:
\begin{equation}\label{axiDdirichlet}
D(k) = -4\left(U(+i0) + |k_2| \int_0^{|k_2|} \frac{U_\rho(\rho + i0)}{\sqrt{k_2^2 - \rho^2}} d\rho\right), \qquad k = ik_2, \quad |k_2| < \rho_0
\end{equation}
and
\begin{equation}\label{axiDneumann}
D(k) = 4\int_{|k_2|}^{\rho_0} \frac{\rho U_\zeta(\rho + i0)}{\sqrt{\rho^2 - k_2^2}} d\rho, \qquad k = ik_2, \quad |k_2| < \rho_0.
\end{equation}
\end{theorem}

It can be verified directly, using Abel transforms, that the integral representation (\ref{axiUfinal}) indeed yields the correct boundary values for $U(\rho, \zeta)$, see appendix \ref{abelapp}.

\section{The Ernst equation}\label{ernstsec}\nequation
\subsection{Lax pair}
The elliptic Ernst equation (\ref{ernst}) admits the Lax pair
\begin{equation}
\begin{cases}\label{ernstlax}
\Phi_z(z, k) = U(z, k) \Phi(z, k),
	\\
\Phi_{\bar{z}}(z, k) =V(z,k) \Phi(z,k),
\end{cases}
\end{equation}
where $z = \rho + i\zeta$, the function $\Phi(z, k)$ is a $2 \times 2$-matrix valued eigenfunction, and the $2\times 2$-matrix valued functions $U$ and $V$ are defined as follows:
$$U = \begin{pmatrix} B & \lambda B \\
\lambda A & A\end{pmatrix}, \qquad 
V =  \begin{pmatrix} \bar{A} & \frac{1}{\lambda} \bar{A} \\
\frac{1}{\lambda} \bar{B} & \bar{B}  
\end{pmatrix},$$
$$A = \frac{f_z}{f + \bar{f}}, \qquad  B = \frac{\bar{f}_z}{f + \bar{f}},$$
with $\lambda = \lambda(z, k)$ given by (\ref{lambdadef}). For each $z$, $\Phi(z, \cdot)$ is a map from the Riemann surface $\mathcal{S}_z$ to the space of $2 \times 2$ matrices. As before, we use the notation $\Phi(z, k)$ to denote $\Phi(\rho, \zeta, k)$, etc.

We can write the Lax pair (\ref{ernstlax}) as
\begin{equation}\label{laxdiffform}  
  d\Phi = W\Phi,
\end{equation}
where $W$ is the one-form
\begin{align}\label{Wdef}  
  W = &Udz + Vd\bar{z}
  	\\ \nonumber
  = &\frac{1}{f + \bar{f}} \begin{pmatrix} \bar{f}_\rho & \frac{1}{2}\left[(\frac{1}{\lambda} + \lambda)\bar{f}_\rho + i(\frac{1}{\lambda} - \lambda)\bar{f}_\zeta\right]	\\
\frac{1}{2}\left[(\frac{1}{\lambda} + \lambda)f_\rho + i(\frac{1}{\lambda} - \lambda)f_\zeta\right]	&	f_\rho
\end{pmatrix}d\rho
	\\ \nonumber
&+ \frac{1}{f + \bar{f}} \begin{pmatrix} \bar{f}_\zeta & \frac{1}{2}\left[(\frac{1}{\lambda} + \lambda)\bar{f}_\zeta - i(\frac{1}{\lambda} - \lambda)\bar{f}_\rho\right]	\\
\frac{1}{2}\left[(\frac{1}{\lambda} + \lambda)f_\zeta - i(\frac{1}{\lambda} - \lambda)f_\rho\right]	&	f_\zeta
\end{pmatrix}d\zeta.
\end{align}
For a $2 \times 2$-matrix $A$, we let $[A]_1$ and $[A]_2$ denote the first and second columns of $A$, respectively.
We define an eigenfunction $\Phi(z, k^\pm)$ as the solution of (\ref{laxdiffform}) which satisfies the initial conditions
\begin{subequations}\label{phiinitial}
\begin{align}\label{phi1initial}
\lim_{z \to i\infty} [\Phi(z, k^-)]_1 = &\begin{pmatrix} 1 \\ 1 \end{pmatrix} \qquad \text{for all $k^-$ on the lower sheet},
	\\ \label{phi2initial}
  \lim_{z \to i\infty} [\Phi(z, k^+)]_2 = &\begin{pmatrix} 1 \\ -1 \end{pmatrix} \qquad \text{for all $k^+$ on the upper sheet}.
\end{align}
\end{subequations}
These initial conditions are convenient because they lead to the following symmetry properties of $\Phi$:
\begin{equation}\label{phisymmetriesk} 
\Phi(z, k^+) = \sigma_3 \Phi(z, k^-) \sigma_1, \qquad 
\Phi(z, k^+) = \sigma_1 \overline{\Phi(z, \bar{k}^+)} \sigma_3.
\end{equation}
Indeed, these symmetries are a consequence of (\ref{phiinitial}), as well as of the following nonlinear analogs of equations (\ref{axiWbarsymmetry}) and (\ref{axiWlift1}):
\begin{align}\label{Wsymmetries}
  W(z, k^+) = \sigma_3W(z, k^-)\sigma_3, \qquad W(z, k^+) = \sigma_1\overline{W(z, \bar{k}^+)}\sigma_1.
\end{align}

\begin{remark}\upshape \label{poleremark}
The Lax pair (\ref{ernstlax}) may have singularities at points where $\text{\upshape Re}\,f = 0$. Physically these points make up the {\it ergospheres} of the spacetime (within these surfaces there can be no static observer with respect to infinity). We will henceforth assume that no ergospheres are present, although we expect our discussion to apply to many cases of physical interest as long as $\text{\upshape Re}\,f > 0$ on the $\zeta$-axis. 
\end{remark}

\subsection{The direct problem}
Assume that $z = \rho + i \zeta$ with $\zeta \geq 0$. Just like for the linear problem, we can express $\Phi(z, k)$ for all $k$ on $\mathcal{S}_z$ by using integration with respect to the contours $\gamma_1$ and $\gamma_2$ in Figure \ref{gamma12contours.pdf}, i.e.
\begin{align}\label{phipositivezeta}
& \Phi(z, k^-) = \lim_{z \to i\infty} \Phi(z, k^-) +  \int_{\gamma_1} (W\Phi)(z', k),
	\\ \nonumber
& \Phi(z, k^+) = \lim_{z \to i\infty} \Phi(z, k^-) + \int_{\gamma_2} (W\Phi)(z', k).
\end{align}
For the linear problem there was a jump across $\Gamma = [-i\rho_0, i\rho_0]$ in the upper sheet. For the nonlinear problem we will find that $\Phi$ jumps both across $\Gamma^+$ and $\Gamma^-$, where $\Gamma^+$ and $\Gamma^-$ denote the coverings of $\Gamma = [-i\rho_0, i\rho_0]$ in the upper and lower sheets, respectively.

The value of $[\Phi(z, k^-)]_1$ as $z \to i\infty$ is fixed by (\ref{phi1initial}), thus an argument similar to that used for the linear problem applied to the contour $\gamma_1$ shows that $[\Phi]_1$ is analytic on the lower sheet. A similar argument applied to the contour $\gamma_2$ shows that $[\Phi]_1$ is analytic on the upper sheet away from the contour $\Gamma^+$.
From the first symmetry in (\ref{phisymmetriesk}), it follows that $[\Phi]_2$ is analytic on the upper sheet and on the lower sheet away from $\Gamma^-$.

\subsection{The inverse problem}\label{inversesubsec}
In order to formulate a RH problem for $\Phi$, we need to determine the `jump matrices' across $\Gamma^+$ and $\Gamma^-$. Suppose that we can express the jump matrix $D$, defined by 
$$\Phi^-(z, k) = \Phi^+(z, k)D(k), \qquad k \in \Gamma^+,$$ 
in terms of the boundary values of $f$ on the disk. Then, the first symmetry in (\ref{phisymmetriesk}) immediately gives the following expression for the jump across $\Gamma^-$:
$$\Phi^-(z, k) = \Phi^+(z, k)\sigma_1 D(k)\sigma_1, \qquad k \in \Gamma^-.$$ 
In the linear problem $D$ was computed {\it directly} in terms of the relevant boundary values. However, for the nonlinear problem it is more convenient to proceed in three steps. In step 1 we express the values of $\Phi$ on the $\zeta$-axis in terms of two spectral functions $F(k)$ and $G(k)$. In step 2 we use the contour $\gamma_1$ to express the jump $D(k)$ in terms of $F(k)$ and $G(k)$. In step 3 we use the contour $\gamma_2$ to relate $F$ and $G$ to the boundary values of $f$ on the disk. Combining steps 2 and 3 we obtain an expression for $D(k)$ in terms of the boundary values of $f$.

\bigskip\noindent
{\bf Step 1.} {\it The values of $\Phi$ on the $\zeta$-axis can be expressed in terms of two spectral functions $F(k)$ and $G(k)$ as
\begin{subequations}\label{phionaxis}
\begin{align} \label{phionaxisa}
\Phi(i\zeta, k^+) =  \begin{pmatrix} \overline{f(i\zeta)} & 1 \\ f(i\zeta) & -1 \end{pmatrix}\begin{pmatrix} F(k) & 0 \\ G(k) & 1 \end{pmatrix}, \qquad \zeta  > 0,\quad k \in \C,
	\\ \label{phionaxisb}
\Phi(i\zeta, k^-) =  \begin{pmatrix} 1 & \overline{f(i\zeta)} \\ 1 & - f(i\zeta) \end{pmatrix}\begin{pmatrix} 1 & G(k) \\ 0 & F(k) \end{pmatrix}, \qquad \zeta  > 0,\quad k \in \C,
	\\ \label{phionaxisc}
\Phi(i\zeta, k^+) = \begin{pmatrix} \overline{f(i\zeta)} & 1 \\ f(i\zeta) & -1 \end{pmatrix}\begin{pmatrix} 1 & G(k) \\ 0 & F(k) \end{pmatrix}, \qquad \zeta  < 0,\quad k \in \C,
	\\ \label{phionaxisd}
\Phi(i\zeta, k^-) = \begin{pmatrix} 1 & \overline{f(i\zeta)} \\ 1 &  -f(i\zeta) \end{pmatrix}\begin{pmatrix} F(k) & 0 \\ G(k) & 1 \end{pmatrix}, \qquad \zeta  < 0,\quad k \in \C.
\end{align}
\end{subequations}
The functions $F(k)$ and $G(k)$ have the following properties:
\begin{itemize}
\item $F$ and $G$ are unique functions of $k \in \C$, i.e. viewed as functions on $\mathcal{S}_z$ they satisfy
\begin{equation}\label{FGunique}
   F(k^+) = F(k^-), \qquad G(k^+) = G(k^-), \qquad k \in \C.
\end{equation}

\item $F(k)$ and $G(k)$ are analytic for $k \in \C \setminus \Gamma$.\footnote{This property holds because of our assumption that $\text{Re}\,f > 0$ on the $\zeta$-axis (the solitonless case). More generally, $F$ and $G$ will have poles which generate solitons. For example, for the Kerr black hole (which is a two-soliton solution), $F(k)$ and $G(k)$ have two poles each on the real axis corresponding to the fact that the real part of $f$ vanishes at two different points on the $\zeta$-axis.}

\item Under the conjugation $k \mapsto \bar{k}$, $F$ andÊ $G$ obey the symmetries
\begin{equation}\label{FGsymm}  
 F(k) = \overline{F(\bar{k})}, \qquad G(k) = -\overline{G(\bar{k})}, \qquad k \in \C.
\end{equation}

\item In the limit $k \to \infty$, 
\begin{equation}\label{FGlimit}  
  F(k) = 1 + O(1/k), \qquad G(k) = O(1/k), \qquad k \to \infty.
\end{equation}
\end{itemize}
}
\proofbegin
For $z = i \zeta$, $\lambda = 1$ for all $k^+$ on the upper sheet and $\lambda = -1$ for all $k^-$ on the lower sheet. Thus
\begin{equation}\label{Wlambda1}
 W(i\zeta, k^+) = \frac{1}{f + \bar{f}} \begin{pmatrix} d\bar{f} & d\bar{f} \\
 df & df\end{pmatrix}.
\end{equation}
We infer that two independent vector-valued solutions of $d\Phi = W(i\zeta, k^+) \Phi$ on the $\zeta$-axis are
\begin{equation}\label{twoindependentplus}  
  \begin{pmatrix} \overline{f(i\zeta)} \\ f(i\zeta) \end{pmatrix}, \qquad \begin{pmatrix} 1 \\ -1 \end{pmatrix}.
\end{equation}  
These solutions together with the initial condition (\ref{phi2initial}) imply that equation (\ref{phionaxisa}) holds for some functions $F(k)$ and $G(k)$. Equation (\ref{phionaxisb}) follows from the first symmetry in (\ref{phisymmetriesk}). The value of $\Phi$ at $z = -i\infty$ is obtained from the value $z = i\infty$ by integrating $W\Phi$ along a large semicircle at infinity. Thus, using (\ref{phionaxisb}) and the fact that $W\Phi$ vanishes for large $z$, we find
\begin{equation}\label{limziinftPhi}
\lim_{z \to -i\infty} \Phi(z, k^+) = \lim_{z \to i\infty} \Phi(z, k^-) = \begin{pmatrix} 1 & 1 \\ 1 & -1\end{pmatrix}\begin{pmatrix} 1 & G(k) \\ 0 & F(k) \end{pmatrix}.
\end{equation}
Using this initial condition together with (\ref{twoindependentplus}), we find (\ref{phionaxisc}). Finally, (\ref{phionaxisd}) follows by symmetry.

By definition, $F$ and $G$ satisfy (\ref{FGunique}). The analyticity properties of $F$ and $G$ follow from the analyticity properties of $\phi(z,k)$. The symmetry properties (\ref{FGsymm}) are a consequence of (\ref{phionaxis}) and of the second symmetry in (\ref{phisymmetriesk}).
In order to prove (\ref{FGlimit}) we note that in the limit $k^+ \to \infty$, equation (\ref{Wlambda1}) holds for all Ê$z$, i.e.
$$\lim_{k \to \infty} W(z, k^+) = \frac{1}{f + \bar{f}}\begin{pmatrix} d\bar{f} & d\bar{f} \\
 df & df\end{pmatrix}.$$
Thus,
$$\lim_{k \to \infty} \Phi(z, k^+) =  \begin{pmatrix} \overline{f(z)} & 1 \\ f(z) & -1 \end{pmatrix}\begin{pmatrix} a & 0  \\ b &  1 \end{pmatrix},$$
for some constants $a, b \in \C$. Evaluating this equation at $z = -i\infty$ and comparing with (\ref{limziinftPhi}), we find that $a = 1$ and $b = 0$. Thus,
$$\Phi(z, k^+) = \begin{pmatrix} \overline{f(z)}  \\ f(z) \end{pmatrix} + O(1/k), \qquad k \to \infty.$$
Comparing this equation with (\ref{phionaxisa}), we find (\ref{FGlimit}).
\proofend

The functions $F(k)$ and $G(k)$ jump across $\Gamma = [-i\rho_0, i\rho_0]$. Let $F^+, G^+$ and $F^-, G^-$ denote the values of $F$ and $G$ for $k$ to the right and left of $\Gamma$, respectively. These jumps are related to $D$, as shown below.

\bigskip\noindent
{\bf Step 2.} {\it The jump $D$ across $\Gamma^+$ is given in terms of $F^\pm$ and $G^\pm$ by
\begin{equation}\label{Dequation1}
  D(k) = \begin{pmatrix} F^+(k) & 0 \\ G^+(k) & 1 \end{pmatrix}^{-1}\begin{pmatrix} F^-(k) & 0 \\ G^-(k) & 1 \end{pmatrix}.
\end{equation}  }
\proofbegin
Integration along the contour $\gamma_1$ with $k$ on the upper sheet gives 
$$\Phi(z, k^+) = \begin{pmatrix} 1 & 1 \\ 1 & -1 \end{pmatrix}\begin{pmatrix} F(k) & 0 \\ G(k) & 1 \end{pmatrix}
+ \int_{\gamma_1} (W\Phi)(z', k^+),$$
where we have used (\ref{phionaxisa}) to determine the initial condition.
Since the branch cut never comes near $\Gamma$ when integrating along $\gamma_1$, the function $W$ in the integrand is continuous across $\Gamma$. Thus the jump $\Delta \Phi := \Phi^+ - \Phi^-$ satisfies
$$\Delta \Phi(z, k^+) = \begin{pmatrix} 1 & 1 \\ 1 & -1 \end{pmatrix}
\left(\begin{pmatrix} F^+(k) & 0 \\ G^+(k) & 1 \end{pmatrix} - \begin{pmatrix} F^-(k) & 0 \\ G^-(k) & 1 \end{pmatrix}\right)
+ \int_{\gamma_1}(W\Delta\Phi)(z', k^+).$$
This shows that $\Delta \Phi$ satisfies the same differential equation as $\Phi^+$, so that there exists a $2\times 2$-matrix valued function $D(k)$ independent of $z$ such that
\begin{equation}\label{Ddef}  
  \Delta \Phi(z, k^+) = \Phi^+(z, k^+)(I - D(k)) \quad \text{i.e.} \quad \Phi^- = \Phi^+D(k).
\end{equation}
Evaluation of (\ref{Ddef}) at $z \to i\infty$, yields the following equation:
$$\begin{pmatrix} 1 & 1 \\ 1 & -1 \end{pmatrix}\left(\begin{pmatrix} F^+(k) & 0 \\ G^+(k) & 1 \end{pmatrix} - \begin{pmatrix} F^-(k) & 0 \\ G^-(k) & 1 \end{pmatrix}\right)
= \begin{pmatrix} 1 & 1 \\ 1 & -1 \end{pmatrix}\begin{pmatrix} F^+(k) & 0 \\ G^+(k) & 1 \end{pmatrix}(I - D(k)).$$
Solving this equation for $D$, we find (\ref{Dequation1}).
\proofend

\begin{remark}\upshape Equation (\ref{Dequation1}) shows that the second column of $D(k)$ is given by $(0, 1)^T$. This is consistent with the observation made earlier that the second column of $\Phi$ has no jump across $\Gamma^+$.
\end{remark}

\bigskip\noindent
{\bf Step 3.} The functions $F^\pm$ and $G^\pm$ satisfy, for $k = ik_2$, $0 < k_2 < \rho_0$, the following equation:
\begin{equation}\label{FGC}
\begin{pmatrix} 1 & G^-(k) \\ 0 & F^-(k) \end{pmatrix} - \begin{pmatrix} 1 & G^+(k) \\ 0 & F^+(k) \end{pmatrix}\begin{pmatrix} F^+(k) & 0 \\ G^+(k) & 1 \end{pmatrix}^{-1}\begin{pmatrix} F^-(k) & 0 \\ G^-(k) & 1 \end{pmatrix}
= \frac{1}{2}\begin{pmatrix} 1 & 1 \\ 1 & - 1 \end{pmatrix} C(k),
\end{equation}  
where the $2\times 2$-matrix valued function $C(k)$ is defined by\footnote{The superscript $-L$ indicates that the function should be evaluated with $k$ lying on the $-$ side of $\Gamma$ and to the left of the branch cut.}
\begin{align}\label{Cdef}
C(k) = \int_{[k_2, \rho_0]} \left[\left((W^R - W^L)\Phi^{-L}\right)(\rho - i0, k^+) - \left((W^R - W^L)\Phi^{-L}\right)(\rho + i0, k^+) \right].
\end{align}
\proofbegin
Consider the expression for $\Phi(z, k^+)$ given in (\ref{phipositivezeta}) by integration along $\gamma_2$.
Let $k^+ = \pm 2\epsilon + ik_2$ and choose the integration contour $\gamma_2$ so that when it passes along the disk, it lies a distance $\epsilon$ above or below the disk (so that $k$ remains on the upper sheet throughout the integration along the disk).
Let $\gamma_2'$ denote the part of $\gamma_2$ which does not run along the disk. Then the $W$ in the integrand is continuous across $\Gamma^+$ during the integration along $\gamma_2'$. It follows that $\Delta \Phi$ satisfies the following equation:
\begin{align*}
\Delta \Phi(z, k^+) = & \lim_{z \to i\infty} \Delta \Phi(z, k^-) 
+ \int_{\gamma_2'} (W\Delta \Phi)(z', k)
	\\
& + \int_{[0, \rho_0]} \bigl[(W\Phi)^{+R}(\rho - i0, k^+) - (W\Phi)^{+R}(\rho + i0, k^+)
  	\\ \nonumber
& - (W\Phi)^{-L}(\rho - i0, k^+) + (W\Phi)^{-L}(\rho + i0, k^+)\bigr].
\end{align*}
The integral along the disk can be written as 
\begin{align*}
  C(k) + \int_{[0, \rho_0]} \bigl[&(W^R\Delta \Phi)(\rho - i0, k^+) - (W^R\Delta \Phi)(\rho + i0, k^+) \bigr],
\end{align*}
where $C(k)$ is given by the right-hand side of (\ref{Cdef}) with the lower integration limit replaced with $0$. Since $W^L = W^R$ for $\rho <k_2$, we see that $C(k)$ can be expressed as in (\ref{Cdef}).
We infer that $\Delta \Phi$ and $\Phi^+$ satisfy the similar equations
$$\Delta \Phi(z, k^+) = \lim_{z \to i\infty} \Delta \Phi(z, k^-) + C(k) + \int_{\gamma_2} (W\Delta \Phi)(z', k), \qquad k \in \Gamma$$
and
$$\Phi^+(z, k^+) = \lim_{z \to i\infty}\Phi^+(z, k^+) + \int_{\gamma_2} (W \Phi^+)(z', k), \qquad k \in \Gamma,$$
where $W$ is $W^R$ for the integrations along the disk. Comparing these equations and using (\ref{phionaxisb}) and (\ref{Ddef}), we find
$$\begin{pmatrix} 1 & 1 \\ 1 & - 1 \end{pmatrix}\left( \begin{pmatrix} 1 & G^+(k) \\ 0 & F^+(k) \end{pmatrix}
- \begin{pmatrix} 1 & G^-(k) \\ 0 & F^-(k) \end{pmatrix}\right) + C(k) 
=\begin{pmatrix} 1 & 1 \\ 1 & - 1 \end{pmatrix} \begin{pmatrix} 1 & G^+(k) \\ 0 & F^+(k) \end{pmatrix}(I - D(k^+)).$$
Using the expression (\ref{Dequation1}) for $D$ and simplifying, we find (\ref{FGC}).
\proofend

\begin{remark}\upshape \label{123remark} 1. The matrix $C(k)$ in (\ref{Cdef}) can be computed in terms of the boundary values of $f$ on the disk via only linear operations.

2. In (\ref{Cdef}) we have assumed that $k_2 > 0$; the jump for $-\rho_0 < k_2 < 0$ can be obtained via the symmetry $D(k) = \sigma_3 \overline{D(\bar{k})} \sigma_3$.

3. Equations (\ref{FGC}) and (\ref{Cdef}) show that, in the case that $W$ is free of singularities, there is no jump at the endpoints of $\Gamma$, i.e. $D(\pm i\rho_0) = I$.
\end{remark}

In order to compute $C(k)$ we need the nonlinear analogs of equations (\ref{axiWrightorabovecut}) and (\ref{axiWLR1}), which express $W$ in the vicinity of the branch cut. On the disk, these equations are
\begin{equation}\label{Wrightorabovecut}
W^{R}(\rho \pm i0, k^+)|_{disk}
= \begin{cases}
\frac{1}{f + \bar{f}} \begin{pmatrix} \bar{f}_\rho & \frac{ik_2 \bar{f}_\rho - \rho \bar{f}_\zeta}{i\sqrt{k_2^2 - \rho^2}} \\
\frac{i k_2 f_\rho - \rho f_\zeta}{i\sqrt{k_2^2 - \rho^2}} & f_\rho 
\end{pmatrix}_{z = \rho \pm i0} d\rho, \qquad \rho < k_2,
	\\
\frac{1}{f + \bar{f}} \begin{pmatrix} \bar{f}_\rho & \frac{ik_2 \bar{f}_\rho - \rho \bar{f}_\zeta}{\sqrt{\rho^2 - k_2^2}} \\
\frac{ik_2 f_\rho - \rho f_\zeta}{\sqrt{\rho^2 - k_2^2}} & f_\rho 
\end{pmatrix}_{z = \rho \pm i0} d\rho, \qquad \rho > k_2
\end{cases}
\end{equation}
and
\begin{equation}\label{WLR}
W^{L}(\rho \pm i0, k^+)
= \begin{cases}
W^{R}(\rho \pm i0, k^+), \qquad \rho < k_2, \\
\sigma_3 W^{R}(\rho \pm i0, k^+) \sigma_3, \qquad \rho > k_2.
\end{cases}
\end{equation}

The above results are summarized in the following proposition.

\begin{proposition}\label{RHprop}
  Let $f(\rho, \zeta)$ satisfy the Ernst equation (\ref{ernst}) in the exterior disk domain $\mathcal{D}$. Suppose that $f$ is asymptotically flat, i.e. $f(\rho, \zeta) \to 1$ as $\rho^2 + \zeta^2 \to \infty$, regular on the rotation axis, i.e. $f_\rho(+0, \zeta) = 0$ for all $\zeta \neq 0$, and that no ergospheres are present, i.e. $\text{\upshape Re} f > 0$ in $\mathcal{D}$. Then the solution $f(\rho, \zeta)$ can be expressed in terms of {\it both} the Dirichlet and the Neumann boundary values on the disk as follows:
\begin{enumerate}
\item Use the boundary values of $f$ to find $W^{R}(\rho \pm i0, k^+)$ and $W^{L}(\rho \pm i0, k^+)$ from equations (\ref{Wrightorabovecut}) and (\ref{WLR}).

\item Find $\Phi^{-L}(\rho \pm i0, k^+)$ for $0 < \rho<\rho_0$ and $0 < k_2 < \rho_0$ by solving the ordinary differential equation
$$\Phi^{-L}_\rho(\rho \pm i0, k^+) = (W\Phi)^{-L}(\rho \pm i0, k^+), \qquad 0 < \rho < \rho_0,$$
together with the initial conditions
$$[\Phi^{-L}(-i0, k^+)]_1 = \begin{pmatrix} \overline{f(-i0)} \\ f(-i0) \end{pmatrix}, \qquad
[\Phi^{-L}(+i0, k^+)]_2 = \begin{pmatrix} 1 \\ -1 \end{pmatrix},$$
as well as the following continuity condition at the tip of the disk:
$$\Phi^{-L}(\rho_0 - i0, k^+) = \Phi^{-L}(\rho_0 + i0, k^+).$$

\item Use the expressions for $W^R$, $W^L$, and $\Phi^{-L}$ to compute $C(k)$ according to (\ref{Cdef}). 

\item Find $D(k)$ from (\ref{Dequation1}) and (\ref{FGC}). The entries of $D$ are rational functions of the entries of $C$.

\item Compute $\Phi$ in terms of the spectral function $D(k)$ via the solution of the following RH problem: 
\begin{itemize}
\item $\Phi(z, k)$ is an analytic function of $k \in \mathcal{S}_z$ away from $k \in \Gamma^\pm$. 

\item Across $\Gamma^+$, $\Phi$ satisfies the jump condition 
$$ \Phi^-(z, k) = \Phi^+(z, k) D(k), \qquad k \in \Gamma^+.$$

\item Across $\Gamma^-$, $\Phi$ satisfies the jump condition 
$$ \Phi^-(z, k) = \Phi^+(z, k) \sigma_1 D(k) \sigma_1, \qquad k \in \Gamma^-.$$

\item As $k \to \infty$, $\Phi$ satisfies
\begin{equation}\label{RHnormalization} 
\lim_{k \to \infty} [\Phi(z, k^-)]_1 = \begin{pmatrix} 1 \\ 1\end{pmatrix}, \qquad \lim_{k \to \infty} [\Phi(z, k^+)]_2 = \begin{pmatrix} 1 \\ -1\end{pmatrix}.
\end{equation}

\item $\Phi$ obeys the symmetries
\begin{equation}\label{PhisymmRH} 
\Phi(z, k^+) = \sigma_3 \Phi(z, k^-) \sigma_1, \qquad 
\Phi(z, k^+) = \sigma_1 \overline{\Phi(z, \bar{k}^+)} \sigma_3.
\end{equation}
\end{itemize}

\item Find $f(z)$ from the equation
\begin{equation}\label{frecover}
  f(z) = \lim_{k \to \infty} (\Phi(z, k^+))_{21}.
\end{equation}  
\end{enumerate}
\end{proposition}

\begin{remark}\upshape
The solution of the RH problem in Proposition \ref{RHprop} is unique, despite the nonstandard form of the normalization condition (\ref{RHnormalization}). Indeed, if $\Phi_1$ and $\Phi_2$ are two solutions of this RH problem, then the functions
$\tilde{\Phi}_j$, $j = 1,2$, defined by
$$\tilde{\Phi}_j(z, k) =\Phi_j(z, \infty^+)^{-1}\Phi_j(z, k), \qquad j = 1,2,$$
satisfy the same jump condition as $\Phi_1$ and $\Phi_2$ together with the standard normalization condition $\tilde{\Phi}_j(z, \infty^+) = I$. Thus $\tilde{\Phi}_1 = \tilde{\Phi}_2$. The equality $\tilde{\Phi}_1(z, \infty^-) = \tilde{\Phi}_2(z, \infty^-)$ together with the first symmetry in (\ref{PhisymmRH}) yield
\begin{equation}\label{PHI12}
  \Phi_1(z, \infty^+)^{-1}\sigma_3\Phi_1(z, \infty^+)\sigma_1 
= \Phi_2(z, \infty^+)^{-1}\sigma_3\Phi_2(z, \infty^+)\sigma_1.
\end{equation}  
Equation (\ref{RHnormalization}) together with second symmetry in (\ref{PhisymmRH}) imply that
\begin{equation}\label{PHIjform}
  \Phi_j(z, \infty^+) = \begin{pmatrix} \overline{f_j(z)} & 1 \\ f_j(z) & -1 \end{pmatrix}, \qquad j = 1,2,
\end{equation}  
where $f_1$ and $f_2$ are complex-valued functions.
Equations (\ref{PHI12}) and (\ref{PHIjform}) imply that $\Phi_1(z, \infty^+) = \Phi_2(z, \infty^+)$. Hence 
$$\Phi_1(z,k) = \Phi_1(z, \infty^+)\tilde{\Phi}_1(z,k) = \Phi_2(z, \infty^+)\tilde{\Phi}_2(z,k) = \Phi_2(z,k).$$
\end{remark}

\section{Equatorial symmetry and the global relation}\label{eqsec}\nequation
Proposition \ref{RHprop} expresses the solution of the Ernst equation in the domain $\mathcal{D}$ via the solution of a RH problem formulated in terms of {\it both} the Dirichlet and Neumann boundary values of $f$ on the disk. Since only a subset of the boundary values can be specified for a well-posed problem, the solution formula is {\it not} yet effective. Recall that for the linear problem, by using equatorial symmetry together with the global relation, we were able to determine the jump data for the RH problem in terms of only either the Dirichlet or the Neumann boundary values. Similarly, for the nonlinear problem, for equatorially symmetric solutions whose boundary values possess a {\it sufficient amount of symmetry} (such boundary values are called {\it linearizable}), the unknown boundary values can be eliminated.

In this section we analyze the consequences of $f$ being equatorially symmetric, i.e. $f(z) = \overline{f(\bar{z})}$, and derive for this case the global relation satisfied by the spectral functions. The linearizable boundary conditions are analyzed in section \ref{linearizablesec}.

\subsection{Equatorial symmetry}\label{eqsubsec}
The nonlinear analog of Proposition \ref{axiWdiskprop} is the following.

\begin{proposition} \label{eqsymmprop}
Assume that $f$ is equatorially symmetric, i.e. $f(z) = \overline{f(\bar{z})}$. Then, for $k \in \Gamma$,
\begin{equation}\label{eqWsymm}
  W^{L}(\rho - i 0, k^+) = \overline{W^{R}(\rho + i0, k^+)}.
\end{equation}
In particular, there exists a $2 \times 2$-matrix valued function $K(k)$, $k \in \Gamma$, independent of $\rho$ such that
\begin{equation}\label{Kdef}
  \Phi^{-L}(\rho - i0, k^+) = \overline{\Phi^{+R}(\rho + i0, k^+)}K(k), \qquad k \in \Gamma.
\end{equation}
The spectral functions $F(k)$ and $G(k)$, introduced in section \ref{inversesubsec}, are related to $K(k)$ as follows:
\begin{equation}\label{Kat0}
  K(k)  = \overline{A_+^{-1}(k)}\sigma_1A_-(k)\sigma_1, \qquad k \in \Gamma,
\end{equation}
where
\begin{equation}\label{Adef}
A(k) := \begin{pmatrix} F(k) & 0 \\ G(k) & 1 \end{pmatrix},
\end{equation}
and $A_\pm$ denote the values of $A$ to the right and left of $\Gamma$ respectively. 
\end{proposition}
\proofbegin
Let $W = W_1 d\rho + W_2 d\zeta$. Then, for any $z$, the definition (\ref{Wdef}) of $W$ yields
\begin{align*}
& \sigma_1\sigma_3W_1(\bar{z}, k^+)\sigma_3\sigma_1 = 
\frac{1}{f(\bar{z}) + \overline{f(\bar{z})}}
\sigma_1\sigma_3
	\\
& \times \begin{pmatrix}
\overline{f_\rho(\bar{z})} & \frac{1}{2}\left[(\frac{1}{\lambda} + \lambda)\overline{f_\rho(\bar{z})} + i(\frac{1}{\lambda} - \lambda)\overline{f_\zeta(\bar{z})} \right]	\\
\frac{1}{2}\left[(\frac{1}{\lambda} + \lambda)f_\rho(\bar{z}) + i(\frac{1}{\lambda} - \lambda)f_\zeta(\bar{z})\right]	&	f_\rho(\bar{z})
\end{pmatrix}_{\lambda = \lambda(\bar{z}, k^+)}
\sigma_3\sigma_1.
\end{align*}
Using the equatorial symmetry, the right-hand side of this equation can be written as
\begin{align*}\label{}
\frac{1}{f(z) + \overline{f(z)}}
\begin{pmatrix}
\overline{f_\rho(z)} 
& \frac{1}{2}\left[-(\frac{1}{\lambda} + \lambda)\overline{f_\rho(z)} + i(\frac{1}{\lambda} - \lambda)\overline{f_\zeta(z)} \right]	\\
\frac{1}{2}\left[-(\frac{1}{\lambda} + \lambda)f_\rho(z) + i(\frac{1}{\lambda} - \lambda)f_\zeta(z)\right]
&	f_\rho(z)
\end{pmatrix}_{\lambda = \lambda(\bar{z}, k^+)}.
\end{align*}
In view of (\ref{lambdazzbarsymmetry}), this expression equals $W_1(z, -k^-)$. Therefore, utilizing the two symmetries in (\ref{Wsymmetries}), we find
$$\sigma_1\sigma_3W_1(\bar{z}, k^+)\sigma_3\sigma_1 = W_1(z, -k^-) = \sigma_3W_1(z, -k^+)\sigma_3 = \sigma_3\sigma_1\overline{W_1(z, -\bar{k}^+)}\sigma_1\sigma_3.$$
This proves that
$$W_1(\bar{z}, k^+) = \overline{W_1(z, -\bar{k}^+)}.$$
Evaluation of this identity for $z = \rho + i0$ on the disk yields (\ref{eqWsymm}). Indeed, if $k^+ = -0 + ik_2$ lies just to the left of $\Gamma$, then $-\bar{k}^+ = 0 + ik_2$ lies just to the right of $\Gamma$. 

It follows from (\ref{eqWsymm}) that $\Phi^{-L}(\rho - i0, k^+)$ and $\overline{\Phi^{+R}(\rho + i0, k^+)}$ satisfy the same differential equation. This establishes the existence of $K(k)$.

Evaluating (\ref{Kdef}) at $\rho = 0$, we find
\begin{equation}\label{preKat0}
   K(k) = \overline{\Phi^{+}(+ i0, k^+)}^{-1} \Phi^{-}(- i0, k^+).
\end{equation}
In view of the axis values (\ref{phionaxis}) of $\Phi$ and the equatorial symmetry, equation (\ref{preKat0}) yields (\ref{Kat0}).
\proofend

\subsection{The global relation}
In this subsection we derive the global relation satisfied by the spectral functions $F(k)$ and $G(k)$.

\begin{proposition}\label{sigma1Kprop}
  Assume that $f(z) = \overline{f(\bar{z})}$ and define $K(k)$, $k \in \Gamma$, by (\ref{Kdef}). Then $K(k)$ satisfies
\begin{equation}\label{sigma1K}
  \sigma_1K(k) = K(k)\sigma_1, \qquad k \in \Gamma.
\end{equation}
\end{proposition}
\proofbegin
For any $z$, it holds that $\Phi(z, k^+) = \Phi(z, k^-)$ for $k = i\bar{z}$ since the two sheets of the Riemann surface coincide at this branch point. Consequently, in view of the first symmetry in (\ref{phisymmetriesk}),
\begin{equation}\label{mathcalFphi}
  \Phi(z, k^+) = \sigma_3 \Phi(z, k^+)\sigma_1, \qquad k = i\bar{z}.
\end{equation}
Evaluating (\ref{mathcalFphi}) for $z$ on the disk, we are able to establish (\ref{sigma1K}). 
Indeed, choosing $z = k_2 + i0$ and $z = k_2 - i0$ in (\ref{mathcalFphi}), and introducing the notation $\Phi_{k_2} := \Phi^+(k_2 + i0, ik_2^+)$, we find
\begin{equation}\label{mathcalFphik2}
  \Phi_{k_2} =  \sigma_3 \Phi_{k_2} \sigma_1
\end{equation}
and
\begin{equation}\label{mathcalFphiminus}
  \Phi^-(k_2 - i0, ik_2^+) = \sigma_3  \Phi^-(k_2 - i0, ik_2^+)  \sigma_1,
\end{equation}
respectively. 
Moreover, evaluating (\ref{Kdef}) at $\rho = k_2$, we find
\begin{equation}\label{Katk2}
  \Phi^{-}(k_2 - i0, k^+) = \bar{\Phi}_{k_2}K(k), \qquad k \in \Gamma.
\end{equation}
Equations (\ref{mathcalFphiminus}) and (\ref{Katk2}) yield
\begin{equation} \label{mathcalFbarphik2K}
  \bar{\Phi}_{k_2}K = \sigma_3 \bar{\Phi}_{k_2}K \sigma_1.
\end{equation} 
Using (\ref{mathcalFphik2}) to replace $\sigma_3 \bar{\Phi}_{k_2}$ with $\bar{\Phi}_{k_2}\sigma_1$, and then premultiplying both sides by $\sigma_1\bar{\Phi}_{k_2}^{-1}$, we find (\ref{sigma1K}).
\proofend

Propositions \ref{eqsymmprop} and \ref{sigma1Kprop} imply the following result.
\begin{proposition}\label{eqprop}
Suppose that $f$ is equatorially symmetric, i.e. $f(z) = \overline{f(\bar{z})}$. 
Then the spectral functions $F(k)$ and $G(k)$ satisfy the following relation, which will be referred to as the global relation:
\begin{equation}\label{eqrelation}
  \overline{A_+(k)}\sigma_1\overline{A_+^{-1}(k)} \sigma_1 = \sigma_1A_-(k) \sigma_1A_-^{-1}(k), \qquad k \in \Gamma,
\end{equation}
where $A(k)$ is defined in terms of $F(k)$ and $G(k)$ by equation (\ref{Adef}). 
\end{proposition}

\begin{remark}\upshape
  The global relation (\ref{eqrelation}) can also be derived by using the nonlinear analog of equation (\ref{globalrelationgeneral}), but the derivation presented here is simpler.
\end{remark}

\section{Linearizable boundary conditions}\label{linearizablesec}\nequation
In Proposition \ref{eqprop} we derived, under the assumption of equatorial symmetry, the global relation (\ref{eqrelation}) satisfied by the spectral functions $F(k)$ and $G(k)$. In this section, we will show that if the corotating potential $f' = f'_0$ is constant on the disk, then the spectral functions $F(k)$ and $G(k)$ satisfy an additional important algebraic relation. 
Furthermore, we will show that these two algebraic relations satisfied by $F(k)$ and $G(k)$ yield an auxiliary RH problem for $F$ and $G$ with jump data given in terms of only the known boundary value $f'_0$. Moreover, this matrix RH problem is diagonalizable and reduces to a scalar RH problem, which can be solved {\it explicitly}. In this way we recover the celebrated Neugebauer-Meinel disk solutions \cite{NM1993}-\cite{NM1995}.

Before considering the consequences of the boundary condition $f' = f_0'$, $f_0'$ constant, on the disk, we first recall the physical origin of the Ernst equation and describe the corotating potential $f'$ (see \cite{MAKNP} for further details). 

\subsection{The Ernst equation}
In canonical Weyl coordinates the exterior gravitational field of a stationarily rotating axisymmetric body is described by the line element
\begin{equation}\label{lineelement}  
  ds^2 = e^{-2U}(e^{2\kappa}(d\rho^2 + d\zeta^2) + \rho^2d\varphi^2) - e^{2U}(dt + a d\varphi)^2,
\end{equation}
where $\rho, \zeta, \varphi$ are cylindrical coordinates, $t$ is the coordinate time, and the metric functions $U, \kappa, a$ depend only on $\rho$ and $\zeta$. In these coordinates, the Einstein field equations consist of the equations
\begin{align}\label{einsteinUeq}
& U_{\rho\rho} + U_{\zeta\zeta} + \frac{1}{\rho}U_\rho = -\frac{e^{4U}}{2\rho^2}\left(a_\rho^2 + a_\zeta^2\right),
	\\ \label{einsteinaeq}
& \left(\rho^{-1}e^{4U}a_\rho\right)_\rho +\left(\rho^{-1}e^{4U}a_\zeta\right)_\zeta = 0,
\end{align}
together with two equations involving $\kappa$. The condition that the metric is regular at the rotation axis implies that
\begin{equation}\label{akonaxis}  
  a \to 0, \, \kappa \to 0 \quad \text{as} \quad \rho \to 0,
\end{equation}  
whereas the condititon that the line element approaches the Minkowski metric at infinity (asymptotic flatness) implies that
\begin{equation}\label{Uakatinfinity}  
  U\to 0, \, a \to 0, \, \kappa \to 0 \quad \text{as} \quad \rho^2 + \zeta^2 \to \infty.
\end{equation}
In view of (\ref{einsteinaeq}), it is possible to introduce a function $b(\rho, \zeta)$ such that
\begin{equation}\label{bdef}  
  a_\rho = \rho e^{-4U}b_\zeta, \qquad a_\zeta = -\rho e^{-4U} b_\rho,
\end{equation}
and
\begin{equation}\label{einsteinbeq}  
  (\rho e^{-4U} b_\rho)_\rho + (\rho e^{-4U} b_\zeta)_\zeta = 0.
\end{equation}
Letting $f = e^{2U} + ib$, equations (\ref{einsteinUeq}) and (\ref{einsteinbeq}) combine into the single Ernst equation (\ref{ernst}). Moreover, as a consequence of the Ernst equation, the compatibility conditions $a_{\rho\zeta} = a_{\zeta\rho}$ and $\kappa_{\rho\zeta} = \kappa_{\zeta\rho}$ are automatically satisfied and the metric functions $a$ and $\kappa$ can be determined by integration of the two equations (\ref{bdef}) and the two field equations for $\kappa$, respectively. Thus, the vacuum Einstein field equations in the stationary axisymmetric case are equivalent to the Ernst equation.

\subsection{Corotating coordinates}\label{corotatingsubsec}
Let us introduce the corotating coordinates $(\rho', \zeta', \varphi', t')$ by\footnote{We will use primes to denote corotating quantities.}
$$\rho' = \rho, \quad \zeta' = \zeta, \quad \varphi' = \varphi - \Omega t, \quad t' = t,$$
where $\Omega$ is the constant angular velocity of the body. In these new coordinates, the metric (\ref{lineelement}) retains its form and the corotating metric functions $U', a', \kappa'$ are related to $U, a, \kappa$ via
\begin{align} \label{U'Urel}
& e^{2U'} = e^{2U}\left[(1 + \Omega a)^2 - \Omega^2\rho^2e^{-4U}\right],
	\\ \nonumber
& (1 - \Omega a')e^{2U'} = (1 + \Omega a)e^{2U},\qquad \kappa' - U' = \kappa - U.
\end{align}
Since the form of the line element is invariant, the field equations retain their form in the corotating system. Thus, we may introduce a corotating Ernst potential $f'$ by $f' = e^{2U'} + i b'$ and the Ernst equation retains its form in the corotating system as well. 

The Lax pair in the corotating system involves an eigenfunction $\Phi'$ and the one-form $W'$ defined by  replacing $f$ with $f'$ in (\ref{Wdef}). The eigenfunction $\Phi'$ is defined as the solution of $d\Phi' = W' \Phi'$ which satisfies the initial conditions (\ref{phiinitial}) with $\Phi$ replaced by $\Phi'$.
It can be verified \cite{MAKNP} that the corotating eigenfunction $\Phi'$ is related to $\Phi$ by
\begin{equation}\label{Lambdadef}
  \Phi'(z, k) = \Lambda(z, k)\Phi(z, k), \qquad k \in \mathcal{S}_z,
\end{equation}
where
\begin{equation}\label{Lambdadef2} 
  \Lambda(z, k) =  (1 + \Omega a)I - \Omega \rho e^{-2U}\sigma_3 + i(k + iz)\Omega e^{-2U}(-\sigma_3 + \lambda(z, k)\sigma_1\sigma_3).
\end{equation}  
Thanks to (\ref{Uakatinfinity}), the relation (\ref{Lambdadef}) is consistent with the requirement that the initial conditions (\ref{phiinitial}) should retain their form for $\Phi'$. 

The spectral analysis of the corotating Lax pair is similar to that of (\ref{ernstlax}), except that $f'$ and $f$ satisfy {\it different} boundary conditions. For example, equation (\ref{U'Urel}) implies that $\text{Re}\,f' \sim 1 - \Omega^2\rho^2$ as $\rho \to \infty$, reflecting the fact that the metric is no longer asymptotically flat in the corotating system. Thus, a given BVP may possess additional symmetries in one of the two coordinate systems. In the next subsection we investigate the consequences of $f'$ being constant along the disk (this corresponds to more complicated boundary values for $f$).

\subsection{$f'$ constant on the disk}\label{fconstsubsec}
We consider the condition $f' = f_0'$, $f_0'$ constant, on the disk. It turns out that the condition of $f'_\rho = 0$ on the disk implies an important relation (see Proposition \ref{fconstprop}) satisfied by the spectral functions $F(k)$ and $G(k)$. The following analysis is conceptually similar to the analysis presented in section \ref{eqsec}; Propositions \ref{fconstsymmprop}-\ref{fconstprop} below are the direct analogs of Propositions \ref{eqsymmprop}-\ref{eqprop}.
The resulting algebraic relation is however independent of the global relation (\ref{eqrelation}). In this subsection we will not assume that $f$ is equatorially symmetric.

\begin{proposition} \label{fconstsymmprop}
Assume that $f'_\rho = 0$ on the disk. Then, for $k \in \Gamma$,
\begin{equation}\label{WRWLbarsymm}  
  W'^R(\rho + i0, k^+) = \sigma_3\sigma_1\overline{W'^L(\rho + i0, k^+)}\sigma_1\sigma_3.
\end{equation}
In particular, there exists a $2\times2$-matrix valued function $Q(k)$, $k \in \Gamma$, independent of $\rho$ such that
\begin{align} \label{Qdef} 
  \overline{\Phi'^{+L}(\rho + i0, k^+)} = \sigma_3 \sigma_1 \Phi'^{+R}(\rho + i0, k^+) Q(k), \qquad k \in \Gamma.
\end{align}
The spectral functions $F(k)$ and $G(k)$ introduced in section \ref{inversesubsec} are related to $Q(k)$ as follows:
\begin{equation}\label{Qat0}
  Q(k) = A_+^{-1}(k)B^{-1}\Lambda^{-1}(k)\sigma_1\sigma_3\overline{\Lambda(k)BA_+(k)}, \qquad k \in \Gamma,
\end{equation}
where $A_+$ is defined in terms of $F$ andÊ $G$ in Proposition \ref{eqsymmprop}, and\footnote{In the presence of equatorial symmetry $\bar{B}$ will be the complex conjugate of $B$.}
\begin{equation}\label{Bdef}
  B := \begin{pmatrix} \overline{f(+ i0)} &  1  \\	 f(+ i0) &   -1 \end{pmatrix}, \qquad \bar{B} := \begin{pmatrix} \overline{f(- i0)} &  1  \\	 f(- i0) &   -1 \end{pmatrix}, \qquad \Lambda(k) := \Lambda(+i0, k^+).
\end{equation}
\end{proposition}
\proofbegin
By the second symmetry in (\ref{Wsymmetries}),
$$\sigma_1\overline{W'^L(\rho + i0, k^+)}\sigma_1 = W'^L(\rho + i0, \bar{k}^+).$$
Thus, for $z = \rho + i0$ on the disk, the corotating analog of the definition (\ref{Wdef}) of $W$ shows that the right-hand side of (\ref{WRWLbarsymm}) is given by
\begin{align}\label{sigma3WLsigma3}
& \sigma_3W'^L(\rho + i0, \bar{k}^+)\sigma_3 
	\\ \nonumber
& = \frac{1}{f' + \bar{f'}} \begin{pmatrix} \bar{f}'_\rho & -\frac{1}{2}\left[(\frac{1}{\lambda} + \lambda)\bar{f}'_\rho + i(\frac{1}{\lambda} - \lambda)\bar{f}'_\zeta\right]	\\
-\frac{1}{2}\left[(\frac{1}{\lambda} + \lambda)f'_\rho + i(\frac{1}{\lambda} - \lambda)f'_\zeta\right]	&	f'_\rho
\end{pmatrix}_{\begin{subarray}{l} z = \rho + i0 \\ \lambda = \lambda^L(z, \bar{k}^+) \end{subarray}} d\rho.
\end{align}
On the other hand,
\begin{align}\label{WRrhoplusi0}
&W'^R(\rho + i0, k^+)
	\\ \nonumber
&= \frac{1}{f' + \bar{f}'} \begin{pmatrix} \bar{f}'_\rho & \frac{1}{2}\left[(\frac{1}{\lambda} + \lambda)\bar{f}'_\rho + i(\frac{1}{\lambda} - \lambda)\bar{f}'_\zeta\right]	\\
\frac{1}{2}\left[(\frac{1}{\lambda} + \lambda)f'_\rho + i(\frac{1}{\lambda} - \lambda)f'_\zeta\right]	&	f'_\rho
\end{pmatrix}_ {\begin{subarray}{l} z = \rho + i0 \\ \lambda = \lambda^R(z, k^+) \end{subarray}} d\rho.
\end{align}
For any $z$,
$$\lambda(z, k^+) = \frac{1}{\lambda\left(z, (-k + 2\zeta)^+\right)}.$$
For $z = \rho + i0$, this equation yields
$$\lambda^R(\rho + i0, k^+) = \frac{1}{\lambda^L(\rho + i0, \bar{k}^+)}.$$
Using this relation to replace $\lambda^L(z, \bar{k}^+)$ with $1/\lambda^R(z, k^+)$ on the right-hand side of (\ref{sigma3WLsigma3}), and then subtracting the resulting equation from equation (\ref{WRrhoplusi0}), we find
\begin{align*}
&W'^R(\rho + i0, k^+) - \sigma_3\sigma_1\overline{W'^L(\rho + i0, k^+)}\sigma_1\sigma_3
	\\
&= \frac{1}{f' + \bar{f}'} \begin{pmatrix} 0 & (\frac{1}{\lambda} + \lambda)\bar{f}'_\rho \\
(\frac{1}{\lambda} + \lambda)f'_\rho 	&	0
\end{pmatrix}_ {\begin{subarray}{l} z = \rho + i0 \\ \lambda = \lambda^R(z, k^+) \end{subarray}} d\rho.
\end{align*}
Setting $f'_\rho = 0$ in this equation, we find (\ref{WRWLbarsymm}).

It follows from (\ref{WRWLbarsymm}) that the functions $\sigma_3 \sigma_1 \Phi'^{+R}(\rho + i0, k^+)$ and $\overline{\Phi'^{+L}(\rho + i0, k^+)}$ satisfy the same differential equation. This establishes the existence of $Q(k)$.

Evaluating (\ref{Qdef}) at $\rho = 0$, we find
\begin{equation}\label{preQat0}
  Q(k) = \Phi'^{+}(+ i0, k^+)^{-1} \sigma_1 \sigma_3 \overline{\Phi'^{+}(+ i0, k^+)}.
\end{equation}
In view of the axis values (\ref{phionaxis}) of $\Phi$ and the definition (\ref{Lambdadef}) of $\Lambda$, equation (\ref{preQat0}) yields (\ref{Qat0}).
\proofend

\begin{proposition}\label{sigma1Qprop}
Assume that $f'_\rho = 0$ on the disk and define $Q(k)$, $k \in \Gamma$, by (\ref{Qdef}). Then $Q(k)$ satisfies
\begin{equation}\label{sigma1Q}
  \sigma_1Q(k) = -Q(k)\sigma_1, \qquad k \in \Gamma.
\end{equation}
\end{proposition}
\proofbegin
Evaluating (\ref{Qdef}) at $\rho = k_2$, we find
$$\bar{\Phi}'_{k_2} = \sigma_3 \sigma_1\Phi'_{k_2} Q(k^+),$$
where $\Phi'_{k_2} := \Phi'^+(k_2 + i0, ik_2^+)$.
The corotating analog of (\ref{mathcalFphik2}) is
$$\Phi'_{k_2} = \sigma_3 \Phi'_{k_2}\sigma_1.$$
The preceding two equations give
\begin{equation}\label{sigmaphik2primeQ}
  -\sigma_1 \Phi'_{k_2}\sigma_1 Q = \sigma_1 \Phi'_{k_2} Q \sigma_1.
\end{equation}
Premultiplying both sides by $-\Phi'^{-1}_{k_2}\sigma_1$, we find (\ref{sigma1Q}).
\proofend

Propositions \ref{fconstsymmprop} and \ref{sigma1Qprop} imply the following result.

\begin{proposition}\label{fconstprop}
Suppose that $f'_\rho = 0$ on the disk. Let $B$, $\bar{B}$, and $\Lambda(k)$ be defined by (\ref{Bdef}).
Then the spectral functions $F(k)$ and $G(k)$ satisfy the relation
\begin{equation*}
  (B^{-1}\Lambda^{-1}\sigma_1\sigma_3\bar{\Lambda}\bar{B})(\bar{A}_+ \sigma_1 \bar{A}_+^{-1}) = - (A_+\sigma_1A_+^{-1})(B^{-1} \Lambda^{-1}\sigma_1\sigma_3 \bar{\Lambda}\bar{B}), \quad k \in \Gamma,
\end{equation*}
where $A_+(k)$ is defined in terms of $F(k)$ and  $G(k)$ in Proposition \ref{eqsymmprop}. 
\end{proposition}

\subsection{The Neugebauer-Meinel disk solutions}
For the BVP denoted by (A) in the introduction, the assumptions of equatorial symmetry and of $f'_\rho = 0$ on the disk are both valid. In this case the spectral functions $F$ and $G$ can be constructed in terms of the known boundary values alone, so that the BVP can be effectively solved. The resulting solutions are the celebrated Neugebauer-Meinel solutions describing rigidly rotating disks of dust cf. \cite{MAKNP}.

Combining Propositions \ref{eqprop} and \ref{fconstprop} we find a $2\times 2$-matrix RH problem for the function $A(k) \sigma_1 A^{-1}(k)$ in the complex $k$-plane with jump across $\Gamma$. In order to express our result in the form presented in \cite{MAKNP}, we formulate this RH problem in terms of the $2\times 2$-matrix valued function $\mathcal{M}(k)$ defined by
\begin{equation}\label{mathcalMdef}
\mathcal{M}(k) = \sigma_3\sigma_1A(k)\sigma_1A^{-1}(k)\sigma_1\sigma_3
= \begin{pmatrix} G(k) &  \frac{G^2(k) - 1}{F(k)} \\ -F(k) & -G(k) \end{pmatrix}.
\end{equation}

\begin{proposition}\label{auxRHprop}
Suppose $f$ is a solution of the BVP denoted by (A) in the introduction. Let $f_0 := f(+i0) = e^{2U_0} + i b_0$ denote the value of $f$ at the origin.
Then the spectral functions $F(k)$ and $G(k)$ are given by
$$F(k) = -\mathcal{M}_{21}(k), \qquad G(k) = \mathcal{M}_{11}(k), \qquad k \in \C,$$
where $\mathcal{M}$ is the unique solution of the following RH problem:
\begin{itemize}
\item \text{$\mathcal{M}(k)$ is analytic for $k \in \C \backslash \Gamma$, $\Gamma = [-i\rho_0, i\rho_0]$.} 

\item  Across $\Gamma$, $\mathcal{M}(k)$ satisfies the jump condition
\begin{equation}\label{auxiliaryRH}  
   \mathcal{S}(k)\mathcal{M}^-(k) = -\mathcal{M}^+(k)\mathcal{S}(k),\qquad k \in \Gamma, 
\end{equation}
where $\mathcal{M}^+$ and $\mathcal{M}^-$ denote the values of $\mathcal{M}$ to the right and left of $\Gamma$, respectively, and $\mathcal{S}(k)$ is defined by
\begin{equation}\label{mathcalSdef}
\mathcal{S}(k) = \begin{pmatrix} f_0 \bar{f_0} - 4\Omega^2k^2 & ib_0 + 2i\Omega k \\ 
ib_0 - 2i\Omega k & -1 \end{pmatrix}, \qquad k \in \Gamma.
\end{equation}

\item $\mathcal{M}$ has the asymptotic behavior
\begin{equation}\label{MtominusI}
  \mathcal{M}(k) = -\sigma_1 + O(1/k), \qquad k \to \infty.
\end{equation}
\end{itemize}
\end{proposition}
\proofbegin
Defining $\mathcal{S}$ by
\begin{equation}\label{mathcalSdef2}
  \mathcal{S} = e^{2U_0} \sigma_3\sigma_1(B^{-1}\Lambda^{-1}\sigma_3\sigma_1\bar{\Lambda}\bar{B})\sigma_3,
\end{equation}
we deduce from Propositions \ref{eqprop} and \ref{fconstprop} that the function $\mathcal{M}$ defined in (\ref{mathcalMdef}) satisfies the jump condition (\ref{auxiliaryRH}). Evaluating (\ref{Lambdadef2}) at $z = 0$ and using (\ref{akonaxis}), we find the following expression for $\Lambda(k)$:
$$\Lambda(k) = I + i k \Omega e^{-2U_0}(\sigma_1 - I)\sigma_3.$$
Substituting this expression for $\Lambda$ together with the expression (\ref{Bdef}) for $B$ into (\ref{mathcalSdef2}), we find that $\mathcal{S}$ can be written as in (\ref{mathcalSdef}).
The asymptotic behavior (\ref{MtominusI}) follows from the properties (\ref{FGlimit}) of $F$ and $G$.
\proofend

\begin{remark}\upshape
1. The RH problem in Proposition \ref{auxRHprop} coincides exactly with the RH problem (2.77) in \cite{MAKNP}. This RH problem can be reduced to a scalar RH problem.
Indeed, the jump condition (\ref{auxiliaryRH}) is of the form
\begin{equation}\label{S1S2RH}  
  \mathcal{S}_1 \mathcal{M}^- = \mathcal{M}^+\mathcal{S}_2, \qquad k \in \Gamma,
\end{equation}
where the invertible matrices $\mathcal{S}_1$ and $\mathcal{S}_2$ are simultaneously diagonalizable, i.e.,
$$\mathcal{S}_j = \mathcal{V} \mathcal{D}_j \mathcal{V}^{-1}, \qquad j = 1, 2,$$
where $\mathcal{D}_1$ and $\mathcal{D}_2$ are diagonal matrices and $\mathcal{V}$ is an invertible matrix. Hence, (\ref{S1S2RH}) can be written as
$$\mathcal{D}_1(\mathcal{V}^{-1} \mathcal{M}\mathcal{V})^- = (\mathcal{V}^{-1}\mathcal{M}\mathcal{V})^+ \mathcal{D}_2.$$
The $(11)$ entry of this equation provides a scalar RH problem which can be solved explicitly. 
In fact, Neugebauer and Meinel were able to combine the auxiliary RH problem of Proposition \ref{auxRHprop} with the main RH problem of Proposition \ref{RHprop} in such a way that the diagonalization and solution of the combined RH problem yields the Ernst potential $f$ directly.

2. For the BVP denoted by (A), the constant value $f_0' = e^{2U_0'} + i b_0'$ of the corotating potential $f'$ on the disk is prescribed. In view of (\ref{U'Urel}), we find $\text{Re}(f_0) = e^{2U_0} = e^{2U_0'}$. On the other hand, the value of the imaginary part of $f_0$, which remains unknown in the formulation of the RH problem (\ref{auxiliaryRH}), disappears in the diagonalization process \cite{MAKNP}.

3. The parameter $b_0'$ can be set to zero without loss of generality, since $b'$ is defined only up to an arbitrary integration constant. Hence the solutions of (A) are parametrized by the three real parameters $U_0, \Omega$, and $\rho_0$. However, we noted in Remark \ref{123remark} that $F$ and $G$ do not jump at the endpoints of the contour $\Gamma$ for a nonsingular solution. Thus, $\mathcal{M}_+ = \mathcal{M}_-$ at the endpoints of $\Gamma$ and we infer from (\ref{auxiliaryRH}) that $\text{Tr}(\mathcal{S}(\pm i \rho_0)) = 0$. This imposes one real condition on the parameters $U_0, \Omega$, and $\rho_0$:
$$|f_0|^2 + 4\Omega^2\rho_0^2 = 1.$$
For physically relevant solutions the Ernst potential $f$ should be nonsingular in all of spacetime. This imposes further restrictions on the parameters $U_0, \Omega$, and $\rho_0$, see \cite{KR1998, KR2005, MAKNP}.
\end{remark}

\appendix
\section{Lax pair singularities} \label{singapp}
\renewcommand{\theequation}{A.\arabic{equation}}\nequation
At the two branch points $k = -iz$ and $k = i\bar{z}$ of the Riemann surface $\mathcal{S}_z$, $\lambda = \infty$ and $\lambda = 0$, respectively; thus the Lax pairs (\ref{axilax}) and (\ref{ernstlax}) have singularities at these points.
In this appendix we analyze in detail the behavior of the eigenfunction $\phi(z, k)$ of the linear problem near the branch points. In particular, we find that $\phi(z, k)$ is nonsingular near these points. Similar statements apply to the eigenfunction $\Phi(z,k)$ of the nonlinear problem.

\begin{proposition}
For a fixed $z = \rho + i\zeta$, $\rho > 0$, the behavior of the map $k \mapsto \phi(z, k):\mathcal{S}_z \to \C$ near $k = -iz$, i.e. near the singularity $\lambda = \infty$, is given by
\begin{equation}\label{phiexpansionapp}
\phi(z, k) = \phi(z, -iz) - 2iU_z(z)\lambda(z, k)(k + iz) + O(k + iz), \qquad k \to -iz,
\end{equation}
where $\phi(z, -iz)$ is a finite number. The behavior of $k \mapsto \phi(z, k)$ near the second branch point $k = i\bar{z}$ follows from (\ref{phiexpansionapp}) and the symmetry (\ref{phiconjugationsymmetry}).
In particular, $\phi(z, \cdot)$ is analytic as a map $\mathcal{S}_z \to \C$ near the branch points.

For a fixed $k$, $\text{\upshape Im}\,k \neq 0$, the behavior of the map $z \mapsto \phi(z, k)$ near the branch point $z = ik$ is given by
\begin{align}\label{phiexpansioninz}
  \phi(z, k) = \phi(ik, k) + 2U_z(ik) \lambda(z, k)(z - ik) + O(|z - ik|^{3/2}), \qquad z \to ik.
\end{align}
\end{proposition}
\proofbegin
Fix $k \in \C$ withÊ $\text{Im}\,k \neq 0$. The Lax pair equations (\ref{axilax}) imply that there exist constants $\epsilon > 0$ and $C > 0$ such that
\begin{align}\label{philaxexpansion}
\begin{cases}
 \phi_z(z, k) = \sqrt{\frac{k - \bar{k}}{k + iz}} U_z(z) + h_1(z, k),
 	\\
 \phi_{\bar{z}}(z, k) = h_2(z, k),
 \end{cases}
  \qquad |z - ik| < \epsilon,
\end{align}	
where the functions $h_1(z, k)$ and $h_2(z,k)$ satisfy
\begin{equation}\label{ghepsilon}
|h_1(z, k)| \leq C \sqrt{|k + iz|}, \quad |h_2(z, k)| \leq C \sqrt{|k + iz|}, \qquad |z - ik| < \epsilon.
\end{equation}
The equations (\ref{philaxexpansion}) show that the value $\phi(ik, k)$ of $\phi(z, k)$ at the branch point $z = ik$, if finite, is given by
\begin{align}\label{phiatbranchpoint}
\phi(ik, k) = \phi(z, k) + \int_{[z, ik]} \left[\left(\sqrt{\frac{k - \bar{k}}{k + iz'}} U_z(z') + h_1(z', k)\right)dz' + h_2(z', k)d\bar{z}'\right], 
	\\ \nonumber
 |z - ik| < \epsilon.
\end{align}
Letting $z' = ik + \rho e^{i\theta}$ with $0 < \rho < r = |z - ik|$, we can write the integral on the right-hand side of (\ref{phiatbranchpoint}) as the following sum of three terms:
$$-\int_0^r \sqrt{\frac{k - \bar{k}}{i\rho e^{i\theta}}} U_z(ik + \rho e^{i\theta})e^{i\theta}d\rho
  -\int_0^r h_1(ik + \rho e^{i\theta}, k)e^{i\theta}d\rho 
-\int_0^r  h_2(ik + \rho e^{i\theta}, k)e^{-i\theta}d\rho.$$
In view of (\ref{ghepsilon}), the integral involving $h_1$ satisfies
$$\left|\int_0^r h_1(ik + \rho e^{i\theta}, k) e^{i\theta}d\rho \right|
\leq \int_0^r  C \sqrt{\rho} d\rho = \frac{2C}{3}r^{3/2}, \qquad 0 < r < \epsilon.$$
A similar estimate holds for the integral involving $h_2$. On the other hand, the integral involving $U_z$ satisfies
\begin{align*}
\int_0^r \sqrt{\frac{k - \bar{k}}{i\rho e^{i\theta}}} U_z(ik + \rho e^{i\theta}) e^{i\theta}d\rho
= & U_z(ik) e^{i\theta} \int_0^r \sqrt{\frac{k - \bar{k}}{i\rho e^{i\theta}}} d\rho + O(r^{3/2})
	\\
=& -2iU_z(ik) \sqrt{i(k - \bar{k})r e^{i\theta}} + O(r^{3/2}), \qquad r \to 0.
\end{align*}
It follows that the integral in (\ref{phiatbranchpoint}) converges, the value $\phi(ik,k)$ at the branch point is finite, and 
\begin{align}\label{phiphiUzsqrt}
\phi(z, k) = \phi(ik, k) - 2iU_z(ik) \sqrt{i(k - \bar{k})(z - ik)} + O(|z - ik|^{3/2}), \qquad z \to ik.
\end{align}
Equation (\ref{phiexpansioninz}) is simply an alternative way of writing this expansion which makes the choice of branches of the square roots more evident (this choice can be fixed by substituting (\ref{phiexpansioninz}) into (\ref{axilax}) and using that $\lambda_z = \frac{-i \lambda}{2(k+ iz)}$).

We now prove (\ref{phiexpansionapp}). For a given $z = \rho + i\zeta$, $\rho > 0$, the map $\phi(z, \cdot)$ from $\mathcal{S}_z$ to $\C$ is analytic in a punctured neighborhood of $k = -iz$ with a possible singularity at $k = -iz$. But equation (\ref{phiphiUzsqrt}) implies that $\phi(z,k)$ is bounded for all $z,k$ near a branch point. Since the function $\phi(z, \cdot)$ is bounded near $k = -iz$, it is in fact analytic in a neighborhood of $k = -iz$. 
For each $z$, we may therefore expand $\phi(z, \cdot)$ in a power series as follows:
$$\phi(z, k) = \phi_0(z) + \phi_1(z) \sqrt{k + iz} + \phi_2(z) (k + iz) + \cdots, \qquad k \to -iz.$$
Substituting this expansion into the first of the Lax pair equations in (\ref{axilax}), the terms of $O(1/\sqrt{k + iz})$ yield
$$\phi_1(z) = -2i\sqrt{-2i\rho}U_z(z).$$
This leads to the expansion
$$\phi(z, k) = \phi(z, -iz) - 2i\sqrt{-2i\rho}U_z(z)\sqrt{k + iz} + O(k + iz), \qquad k \to -iz.$$
Equation (\ref{phiexpansionapp}) is an alternative way of writing this equation which makes the choice of branches more evident.

\begin{remark}\upshape
The integral representation (\ref{axiRHsolution}) for $\phi(z,k)$ is consistent with the expansions (\ref{phiexpansionapp}) and (\ref{phiexpansioninz}). This can be checked by using the following representation for $U_z$ obtained from (\ref{preaxiUrep}),
$$U_z(z) =- \frac{1}{8\pi}\int_\Gamma \frac{D(k)}{\lambda(z, k)(k + iz)^2} dk.$$
\end{remark}

\section{Abel transforms} \label{abelapp}
\renewcommand{\theequation}{B.\arabic{equation}}\nequation
In this appendix we use Abel transforms to verify explicitly that the integral representation of the solutionÊ $U(\rho, \zeta)$ of the axisymmetric Laplace equation given in Theorem \ref{axith} indeed yields the correct boundary values. 

Let $U(\rho, \zeta)$ be defined by
\begin{equation}\label{axiUfinalapp}
U(\rho, \zeta) = -\frac{1}{4\pi i}\int_{-i\rho_0}^{i\rho_0} \frac{D(k)}{\sqrt{(k - \zeta)^2 + \rho^2}} dk,
\end{equation}
where the branch with positive real part is chosen for the square root, and $D(k)$ is given by (\ref{axiDdirichlet}) with $U(\rho + i0)$ replaced by $U_0(\rho)$, i.e.
\begin{equation}\label{Ddirichletapp} 
 D(k) = -4\left(U_0(0) + |k_2| \int_0^{|k_2|} \frac{U_{0\rho}(\rho)}{\sqrt{k_2^2 - \rho^2}} d\rho\right), \qquad k = ik_2, \quad |k_2| < \rho_0.
\end{equation}
We need to show that $U = U_0$ on the disk.

For $z = \rho + i 0$ just above the disk, we find
\begin{equation}\label{sqrtapp}
\sqrt{(k - \zeta)^2 + \rho^2}\bigl|_{\zeta = +0} = \begin{cases} -i \sqrt{k_2^2 - \rho^2},Ê\qquad \rho < k_2 < \rho_0, \\
\sqrt{\rho^2 - k_2^2}, \qquad |k_2| < \rho, \\
i\sqrt{k_2^2 - \rho^2}, \qquad -\rho_0 < k_2 < -\rho.
\end{cases}
\end{equation}
Equations (\ref{axiUfinalapp}) and (\ref{sqrtapp}) imply that
\begin{equation}\label{axiUondiskapp}
U(\rho, +0) = -\frac{1}{2\pi}\int_{0}^{\rho} \frac{dk_2}{\sqrt{\rho^2 - k_2^2}} D(ik_2), \qquad 0 < \rho <  \rho_0.
\end{equation}
Defining the Abel transform $\hat{h}(k_2)$ of a function $h(\rho)$ by
\begin{equation}\label{abel}  
\hat{h}(k_2) = \frac{1}{\pi} \left(\frac{h(0)}{k_2} + \int_0^{k_2} \frac{h_{\rho}(\rho) d\rho}{\sqrt{k_2^2 - \rho^2}}\right)
\end{equation} 
equation (\ref{Ddirichletapp}) can be written as
\begin{equation}  \label{Ddirichletabel}
  D(ik_2) = -4\pi|k_2| \hat{U}_0(|k_2|).
\end{equation}
Substituting (\ref{Ddirichletabel}) into (\ref{axiUondiskapp}) and using that the inverse of (\ref{abel}) is given by
\begin{equation*}\label{}  
  h(\rho) = 2 \int_0^\rho \frac{\hat{h}(k_2) k_2 dk_2}{\sqrt{\rho^2 - k_2^2}},
\end{equation*}
we infer that indeed $U(\rho, +0) = U_0(\rho)$.

In order to verify the integral representation in terms of the Neumann boundary values, we let $D(k)$ be given by (\ref{axiDneumann}) with $U_\zeta(\rho + i0)$ replaced by $U_1(\rho)$, i.e.
\begin{equation}\label{axiDneumannapp}
D(k) = 4\int_{|k_2|}^{\rho_0} \frac{\rho U_1(\rho)}{\sqrt{\rho^2 - k_2^2}} d\rho, \qquad k = ik_2, \quad |k_2| < \rho_0.
\end{equation}
We need to show that $U_\zeta(\rho, +0) = U_1(\rho)$.
We define the following slight variation of the Abel transform (\ref{abel}):
\begin{equation}\label{abel2}  
\tilde{h}(k_2) = -\frac{1}{\pi k_2}\frac{d}{dk_2}\int_{k_2}^{\rho_0} \frac{h(\rho)\rho d\rho}{\sqrt{\rho^2 - k_2^2}},
\end{equation}
whose inverse is given by
\begin{equation}\label{inverseabel2}  
  h(\rho) = 2 \int_{\rho}^{\rho_0} \frac{\tilde{h}(k_2) k_2 dk_2}{\sqrt{k_2^2 - \rho^2}}.
\end{equation}
Equation (\ref{axiDneumannapp}) yields
\begin{equation}\label{dDdk2}
\frac{d}{dk_2}D(ik_2) = -4\pi k_2 \widetilde{U_\zeta}(k_2), \qquad   \rho < k_2 < \rho_0.
\end{equation}
On the other hand, equation (\ref{axiUfinalapp}) yields
\begin{equation}\label{Uzetaapp2}
U_\zeta(\rho, +0) = - \frac{1}{4\pi}\frac{d}{d\zeta}\biggl|_{\zeta = 0^+} \int_{-\rho_0}^{\rho_0} \frac{D(ik_2) dk_2}{\sqrt{(ik_2 - \zeta)^2 + \rho^2}}.
\end{equation}
The function $g(\rho, \zeta, k_2)$ defined by
$$g(\rho, \zeta, k_2) = i\log\left(k_2 + i\zeta + i\sqrt{(ik_2 - \zeta)^2 + \rho^2}\right),$$
satisfies
$$\frac{dg}{dk_2} = \frac{1}{\sqrt{(ik_2 - \zeta)^2 + \rho^2}}.$$
Thus, integrating by parts in (\ref{Uzetaapp2}) and using that $D(\pm i\rho_0) = 0$, we find
$$U_\zeta(\rho, +0) = \frac{1}{4\pi}\frac{d}{d\zeta}\biggl|_{\zeta = 0^+} \int_{-\rho_0}^{\rho_0} g(\rho, \zeta, k_2) \frac{d}{dk_2} D(ik_2) dk_2.$$
We can now interchange the orders of differentiation and integration. The identity
$$\frac{dg}{d\zeta} = i \frac{dg}{dk_2} =   \frac{i}{\sqrt{(ik_2 - \zeta)^2 + \rho^2}},$$
implies that
$$U_\zeta(\rho, +0) =  \frac{1}{4\pi} \int_{-\rho_0}^{\rho_0} \frac{i}{\sqrt{(ik_2 - \zeta)^2 + \rho^2}\bigl|_{\zeta = 0^+} }\frac{d}{dk_2} D(ik_2) dk_2.$$
Using (\ref{sqrtapp}) together with the fact that $\frac{d}{dk_2} D(ik_2)$ is an odd function, we infer that
\begin{align}\label{Uzetaapp3}
U_\zeta(\rho, +0) = -\frac{1}{2\pi} \int_{\rho}^{\rho_0} \frac{1}{\sqrt{k_2^2 - \rho^2} }\frac{d}{dk_2} D(ik_2) dk_2.
\end{align}
Substituting (\ref{dDdk2}) into (\ref{Uzetaapp3}) and using (\ref{inverseabel2}), we find that indeed $U_\zeta(\rho, 0) = U_{1}(\rho)$.

 \bigskip
\noindent
{\bf Acknowledgement} {\it The authors acknowledge support from a Marie Curie Intra-European Fellowship and the Guggenheim foundation.}

\bibliography{is}

\end{document}